\documentclass[aps,prb,twocolumn,reprint,amsmath,amssymb,floatfix,eqsecnum]{revtex4}
\usepackage{graphicx}
\usepackage{amssymb,bbm,amsmath,braket,indentfirst,bm}
\allowdisplaybreaks 
\usepackage[pdftex]{color} %Titus 
\usepackage{hyperref}
\usepackage{color}
\usepackage{ulem} % to strike things out \normalem % usual emph
\usepackage[paperwidth=8.5in,paperheight=11in,centering,hmargin=2cm,vmargin=2.5cm]{geometry} % fix margins to be regular all around

\newcommand{\expect}[1]{\ensuremath \left\langle #1 \right\rangle}

\newcommand{\abs}[1]{\ensuremath \left|#1\right|}

\newcommand{\pri}[1]{\ensuremath #1^{\prime}}

\newcommand{\der}[2]{\ensuremath \frac{\mathrm{d}#1}{\mathrm{d}#2}} %1st derivative
\renewcommand\({\ensuremath \left(} \renewcommand\){\ensuremath
  \right)} \renewcommand\[{\ensuremath \left[}
  \renewcommand\]{\ensuremath \right]}
\def\:={\,\raisebox{0.85pt}{.}\hspace{-2.78pt}\raisebox{2.85pt}{.}\!\!=\,}
\def\=:{\,=\!\!\raisebox{0.85pt}{.}\hspace{-2.78pt}\raisebox{2.85pt}{.}\,}

\begin{document}

\title{Floquet systems coupled to particle reservoirs}

\author{Thomas~Iadecola} \affiliation{Physics Department, Boston
  University, Boston, Massachusetts 02215, USA}

\author{Claudio~Chamon} \affiliation{Physics Department, Boston
  University, Boston, Massachusetts 02215, USA}

\date{\today}

\begin{abstract}
Open quantum systems, when driven by a periodic field, can relax to effective statistical ensembles that resemble their equilibrium counterparts.  We consider a class of problems in which a periodically-driven quantum system is allowed to exchange both energy and particles with a thermal reservoir.  We demonstrate that, even for noninteracting systems, effective equilibration to the grand canonical ensemble requires both fine tuning the system-bath coupling and selecting a sufficiently simple driving protocol.  We study a tractable subclass of these problems in which the long-time steady state of the system can be determined analytically, and demonstrate that the system effectively thermalizes with fine tuning, but does not thermalize for general values of the system-bath couplings.  When the driven system does not thermalize, it supports a tunable persistent current in the steady state without external bias.  We compute this current analytically for two examples of interest: 1) a driven double quantum dot, where the current is interpreted as a DC electrical current, and 2) driven Dirac fermions in graphene, where it is interpreted as a valley current.
\end{abstract}

\maketitle

\section{Introduction}

Since Floquet's theorem~\cite{floquet} for linear ordinary differential equations with time-periodic coefficients was introduced to quantum mechanics by Shirley~\cite{shirley} and Sambe,~\cite{sambe} Floquet theory has found extensive applications in the study of cold atomic gases, where it is used to design properties of many-body Hamiltonians.~\cite{eckardt,aidelsburger,jotzu,bukov1,bukov2}  These systems can be prepared in such a way that they are very well isolated from the environment.  There is also, however, great interest in using periodic driving to design transport and band-structure properties of solid state systems, such as semiconductors and semiconductor heterostructures,~\cite{lindner,goldstein} graphene,~\cite{oka,kitagawa,iadecola1,grushin,iadecola3} and topological insulator surface states.~\cite{fregoso,wang}  However, in these systems, which realistically must be coupled to a reservoir (e.g.~phonons or a substrate), the band structure of the Floquet effective Hamiltonian is not sufficient to predict their steady-state electronic properties at long times.  Instead, the coupling to the reservoir must be taken into account, so that the density matrix of the system at long times can be determined accurately.

The statistical mechanics of periodically-driven open quantum systems has been of great interest and concern to condensed matter physicists over the past several decades, and has been relevant to problems in atomic physics\cite{blumel} and quantum transport.~\cite{kohler_review,camalet,lehmann,kohler04}  In these contexts, the thermal reservoirs, such as quantized radiation fields and metallic leads, help these systems relax to periodic steady states at long times.  The study and characterization of these steady states has become an active subfield of physics in its own right, known as periodic thermodynamics.~\cite{kohler,breuer,kohn,hone,ketzmerick,langemeyer}

There are cases where interactions with an external reservoir can bring about the relaxation of a periodically-driven system to a steady state in which the occupations of the Floquet states follow an equilibrium distribution akin to the canonical ensemble.  In these systems, one recovers the physics of undriven systems, with the Floquet quasienergies playing the role once played by the conventional energies at equilibrium.  When this occurs, we will refer to the system as ``effectively thermalized."  Effectively-thermalized states have been investigated theoretically in the periodically-forced quantum harmonic oscillator,~\cite{breuer} as well as in driven Dirac fermion systems.~\cite{iadecola1,iadecola2,iadecola3}  More recently, the conditions under which a periodically-driven quantum system coupled to a reservoir effectively thermalizes have been investigated for system-bath couplings that preserve the number of particles in the system.~\cite{shirai,liu}  There is also numerical evidence that certain systems relax to such a distribution in the high-frequency limit.~\cite{liu}  

However, it is known on general grounds~\cite{kohn,hone} that generic periodically-driven quantum systems do not thermalize to an effective canonical statistical ensemble when coupled to reservoirs in such a way that the number of particles in the system is conserved.  Indeed, there are many examples of simple Floquet systems coupled to simple reservoirs where effective thermalization does not occur.~\cite{hone,langemeyer,dehghani,d'alessio}  In this work, we show that the same is true when the system is also allowed to exchange particles with the bath---generic periodically-driven quantum systems do not thermalize to an effective grand canonical ensemble when coupled to a particle reservoir.  We are therefore led to the following point of view: while it is important to understand the limits in which open Floquet systems reach an effective thermal equilibrium with a reservoir, the vast majority of such systems do not equilibrate in this way.  Consequently, one should elevate the systems that do {\it not} reach an effective thermal equilibrium to a similar level of importance; the steady states of these systems could exhibit novel phases that are inaccessible at equilibrium, and may possess useful tunable characteristics.

To this end, the structure of the paper is as follows.  We begin by introducing the general setup that we wish to consider, namely cases where a periodically-driven quantum system is coupled to a reservoir with which it can exchange both energy and particles.  In addition to the more general motivation that such a setup allows us to investigate the possibility of a Floquet equivalent of the grand canonical ensemble, we also note that problems of this type are ubiquitous in driven quantum transport, where the leads play the role of a particle reservoir.  We move on to discuss conditions under which such a system thermalizes to an effective grand-canonical quasienergy distribution, and suggest a criterion for this to occur in noninteracting systems for a common class of system-bath couplings.  This criterion illustrates that these systems can only equilibrate to the effective grand canonical ensemble when both the system-bath coupling and the driving protocol are fine-tuned.  

\begin{figure}[t]
\includegraphics[width=.475\textwidth]{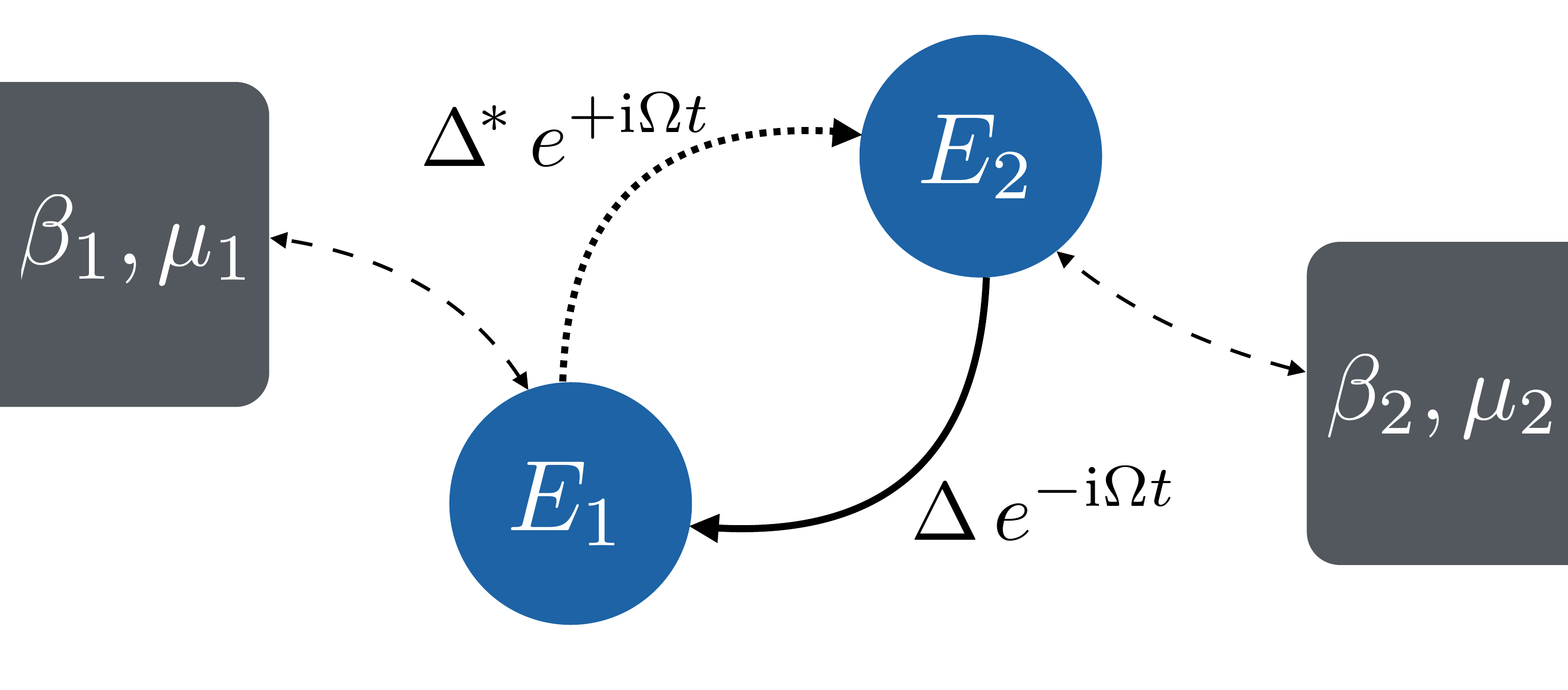}
\caption{(Color online) The driven double quantum dot setup discussed in Sec.~\ref{double dot example}.  The two single-level quantum dots at energies $E_{1,2}$ are weakly coupled to two reservoirs at inverse temperatures $\beta_{1,2}$ and chemical potentials $\mu_{1,2}$.  Transitions between the two dots occur preferentially via absorption (dotted line) or emission (solid line) of energy $\Omega$.
\label{fig:double_dot}
        }
\end{figure}

\begin{figure}[t]
\includegraphics[width=.475\textwidth]{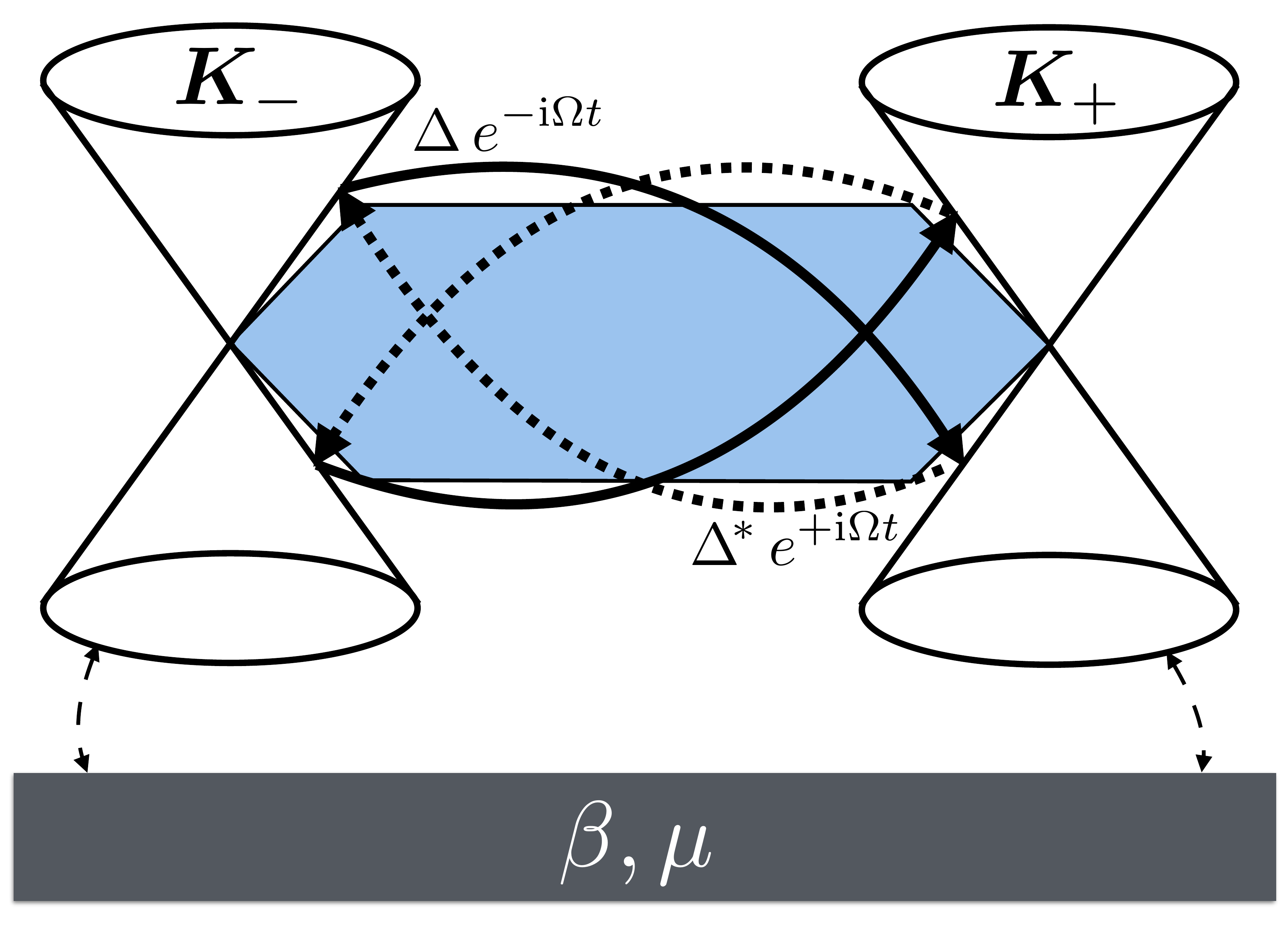}
\caption{(Color online) Graphene in the presence of a rotating Kekul\'e mass term (c.f.~Sec.~\ref{graphene example}).  The two inequivalent Dirac cones located at opposite corners of the Brillouin zone (blue hexagon) are coupled to a common reservoir at inverse temperature $\beta$ and chemical potential $\mu$.  The driving excites transitions from one cone to the other, which occur preferentially via either absorption (dotted lines) or emission (solid lines) of energy $\Omega$.
\label{fig:graphene}
        }
\end{figure}

We then move on to a detailed study of a particular class of models in which two species of fermions are driven in such a way that transitions between the two species are excited via either absorption or emission.  This class of models is exceptionally simple in that the steady state of the system at long times can be calculated analytically within the Born-Markov approximation.  We calculate the steady-state populations and coherences for this class of models and find that the system only effectively thermalizes when the system-bath coupling is fine-tuned to a critical line in parameter space.  When the system does not effectively thermalize, an interesting steady state emerges in which a persistent current arises without external bias.  This current can be tuned by varying the driving parameters, and it depends on the bath density of states.  We analyze the nonthermal distribution of occupations and the persistent currents for two examples: 1) a driven double quantum dot (Fig.~\ref{fig:double_dot}) and 2) driven Dirac fermions in graphene (Fig.~\ref{fig:graphene}).  The persistent current is interpreted as a DC electric current in the former case and a valley current in the latter.  The possibility of engineering such persistent currents in other solid-state systems presents an enticing direction for future work.

\subsection{General model}
Let us consider a periodically driven quantum system coupled to a thermal reservoir.  Generically, systems of this type are modeled by a time-dependent Hamiltonian of the form
\begin{align}\label{H total}
H(t)&=H_{\rm S}(t)+H_{\rm SB}+H_{\rm B},
\end{align}
where $H_{\rm S}(t)$ describes the system of interest, and depends explicitly on time due to the driving.  $H_{\rm S}(t)$ is assumed to be periodic in time with period $T=2\pi/\Omega$, where $\Omega$ is the driving frequency.  $H_{\rm B}$ describes the thermal reservoir (usually a noninteracting model with many more degrees of freedom than the system) at inverse temperature $\beta$ and chemical potential $\mu$, and $H_{\rm SB}$ describes the interaction between the system and reservoir.

For the purposes of our discussion, it will be important to distinguish between two different types of system-bath couplings $H_{\rm SB}$.  In particular, we will draw a distinction between system-bath couplings that do and do not allow for particle transport between the system and the reservoir.  An example of the latter type of system-bath coupling is
\begin{align}
H_{\rm SB}=\sum_{\bm k,\alpha,\beta,n} g^n_{\alpha\beta}\ c^\dagger_{\bm k,\alpha} c_{\bm k,\beta}\(b^\dagger_{\bm k,n}+b_{\bm k,n}\),
\end{align}
where the operator $c^\dagger_{\bm k,\alpha}$ creates a particle with momentum $\bm k$ in the system and the operator $b^\dagger_{\bm k,\alpha}$ creates a particle with momentum $\bm k$ in the bath.  The indices $\alpha$ and $\beta$ label degrees of freedom in the Hilbert space of the system, while the index $n$ labels degrees of freedom in the Hilbert space of the bath.  Since the bath degrees of freedom couple to an operator that preserves the total number of particles in the system, the above choice of $H_{\rm SB}$ allows the system to exchange energy with the bath while keeping the total particle number fixed.  An example of the other type of system-bath coupling is
\begin{align}\label{grand canonical example}
H_{\rm SB}=\sum_{\bm k,\alpha,n} \(g^n_{\alpha}\ c^\dagger_{\bm k,\alpha} b_{\bm k,n}+g^{n\, *}_\alpha\ b^\dagger_{\bm k,n} c_{\bm k,\alpha}\).
\end{align}
In this case, the system is able to exchange both particles and energy with the bath, as the bath no longer couples to a bilinear system operator.  However, the total number of particles in the full closed system is still conserved.  

In this work, we will be primarily interested in systems of noninteracting fermions placed in contact with fermionic reservoirs via a coupling of the form \eqref{grand canonical example}.  However, all analyses carried out in this paper can be trivially modified to apply to bosonic systems in contact with bosonic reservoirs.

\subsection{Conditions for equilibration with the reservoir}
It was noted by Breuer et al.,\cite{breuer} Hone et al.~\cite{hone} and by us\cite{iadecola2} that, under certain conditions, it is possible for a periodically driven quantum system to relax to a steady state that resembles the equilibrium distribution of a time-independent system.  These conditions have been sharpened in more recent work.~\cite{shirai,liu}  We briefly review these conditions here, and refer the reader to the original references for more details.  First, let us recall that, according to Floquet's theorem, a solution $\ket{\Psi_\alpha(t)}$ of the time-dependent Schr\"odinger equation $\[H_{\rm S}(t)-i\partial_t\]\ket{\Psi_\alpha(t)}=0$ can be written in a particular form:
\begin{align}\label{floquet theorem}
\ket{\Psi_\alpha(t)} = e^{-i\varepsilon_\alpha t}\ket{\Phi_\alpha(t)},
\end{align}
where the ``Floquet state" $\ket{\Phi_\alpha(t)}=\ket{\Phi_\alpha(t+T)}$ shares the time-periodicity of $H_{\rm S}(t)=H_{\rm S}(t+T)$.  Substituting \eqref{floquet theorem} into the time-dependent Schr\"odinger equation yields an eigenvalue problem for the quasi-energy $\varepsilon_\alpha$:
\begin{align}\label{floquet eigenvalue problem}
\[H_{\rm S}(t)-i\partial_t\]\ket{\Phi_\alpha(t)}=\varepsilon_\alpha\ket{\Phi_\alpha(t)}.
\end{align}
Since the states $\ket{\Phi_\alpha(t)}$ are periodic, we substitute $\ket{\Phi_\alpha(t)}=\sum_{n=-\infty}^\infty e^{-in\Omega t}\ket{\Phi_\alpha^n}$, into \eqref{floquet eigenvalue problem} to obtain
\begin{align}
\sum_{n=-\infty}^\infty \[H_{{\rm S}, mn}-m\Omega\; \delta_{mn}\]\ket{\Phi_\alpha^n}=\varepsilon_\alpha\ket{\Phi_\alpha^m},
\end{align}
where $H_{{\rm S},mn}=\frac{1}{T}\int_0^Tdt\; e^{i(m-n)\Omega t}\, H_{\rm S}(t)$.  The eigenvalue problem then acquires a matrix structure in the space of Fourier harmonics:
\begin{subequations}\label{matrix form}
\begin{align}
\mathcal H\Phi_\alpha &= \varepsilon_\alpha\Phi_\alpha,\\
\mathcal H&=\begin{pmatrix}
\ddots& & & &\\
\cdots&H_0-\Omega\, \mathbbm 1&H_{+1}&H_{+2}&\cdots\\
\cdots&H_{-1}&H_0&H_{+1}&\cdots\\
\cdots&H_{-2}&H_{-1}&H_0+\Omega\, \mathbbm 1&\cdots\\
& & & &\ddots
\end{pmatrix}\\
\Phi_\alpha&=\begin{pmatrix}&\dots&\ket{\Phi_\alpha^{1}}&\ket{\Phi_\alpha^0}&\ket{\Phi_\alpha^{-1}}&\dots&\end{pmatrix}^\textsf{T},
\end{align}
where we have defined
\begin{align}
H_{n}&=\frac{1}{T}\int_0^T dt\; e^{-\,  \mathrm i\, n \Omega\, t}\, H_{\rm S}(t)\\
H_{\rm S}(t)&=\sum^{\infty}_{n=-\infty}e^{+\mathrm i\, n\Omega\, t}\, H_{n}.
\end{align}
\end{subequations}

There is an ambiguity in the definition of the quasienergies $\varepsilon_\alpha$, as the definition \eqref{floquet theorem} is invariant under shifts $\varepsilon_\alpha\to \varepsilon_\alpha+m_\alpha\Omega$, where $m_\alpha$ is an integer.  If the driving frequency $\Omega$ is of roughly the same order as other energy scales in the problem, this can lead to an ambiguity in the ordering of the quasienergies as well.  This ambiguity can be remedied (c.f.~Refs.~\onlinecite{iadecola2} and \onlinecite{liu}) by viewing the quasienergies as the eigenvalues of an effective time-independent Hamiltonian, which is defined as follows.  Formally, one can always block-diagonalize the infinite-dimensional Hermitian matrix $\mathcal H$ by a unitary transformation $\mathcal U$ such that
\begin{subequations}
\begin{align}
\mathcal U^\dagger\, \mathcal H\, \mathcal U 
&=
\begin{pmatrix}
& \ddots & & & &&\\
&\cdots&H_{\rm eff}-\omega&0&0&\cdots&\\
&\cdots&0&H_{\rm eff}&0&\cdots&\\
&\cdots&0&0&H_{\rm eff}+\omega&\cdots&\\
&&&&&\ddots &
\end{pmatrix},\\
\mathcal U
&=
\begin{pmatrix}
\ddots& & & &\\
\cdots&U_0&U_{+1}&U_{+2}&\cdots\\
\cdots&U_{-1}&U_0&U_{+1}&\cdots\\
\cdots&U_{-2}&U_{-1}&U_0&\cdots\\
& & & &\ddots
\end{pmatrix}.
\end{align}
\end{subequations}
The time-independent effective Hamiltonian is then
\begin{subequations}
\begin{align}\label{h effective}
H_{\rm eff}&= U^\dagger(t)\, H_{\rm S}(t)\, U(t)-\mathrm i\, U^\dagger(t)\, \partial_t\, U(t),
\end{align}
where the time-dependent unitary transformation $U(t)$ is defined by
\begin{align}
U(t)&=\sum^{\infty}_{n=-\infty}e^{+\mathrm i\, n\Omega\, t}\, U_{n}.
\end{align}
\end{subequations}
Observe that, by construction, $U(t)$ is periodic with the same period $T=2\pi/\Omega$ as the driving in $H_{\rm S}(t)$.  [Indeed, the existence of this time-periodic operator that transforms $H_{\rm S}(t)$ into $H_{\rm eff}$ is guaranteed by Floquet's theorem, as it is identical to the ``kick operator" that appears in the factorization of the time-evolution operator.\cite{goldman,bukov1}]  The shift symmetry $\varepsilon_\alpha\to \varepsilon_\alpha+m_\alpha\Omega$ ($m_\alpha\in\mathbb Z$) then manifests itself as a gauge symmetry of $H_{\rm eff}$ under the discrete set of time-dependent transformations of the form
\begin{align}
H_{\rm eff}\to e^{-\mathrm i\, M\Omega\, t}\, H_{\rm eff}\, e^{+\mathrm i\, M\Omega\, t}-\mathrm i\, (e^{-\mathrm i\, M\Omega\, t})\partial_t(e^{+\mathrm i\, M\Omega\, t}),
\end{align}
with diagonal matrices $M_{\alpha\beta}=m_\alpha\, \delta_{\alpha\beta}$, which do not alter the time-periodicity of the Floquet states.

While the transformation $U(t)$ that brings the time-periodic Hamiltonian $H_{\rm S}(t)$ into the time-independent form of Eq.~\eqref{h effective} always exists, it is usually not simple to compute, and, in most cases, can only be obtained perturbatively.~\cite{goldman,verdeny,bukov1}   There also exist classes of examples in which it can be computed exactly.\cite{rabi,iadecola1,iadecola2,iadecola3,langemeyer,bukov1}  However, one can nevertheless use the existence of such a transformation to point out conditions under which one could expect relaxation to an effective equilibrium ensemble.  Suppose, for example, that the (time-independent) system-bath coupling takes the generic form
\begin{align}
H_{\rm SB}=g\, S\, B,
\end{align}
where $S$ consists of some Hermitian linear combination of system operators and $B$ consists of some Hermitian linear combination of bath operators, and $g$ is a real constant that parameterizes the strength of the system-bath coupling.  Then, one expects that if there exists a basis in which the system Hamiltonian and the system-bath coupling are simultaneously time-independent, i.e. if
\begin{align}
S_{\rm eff}=U^\dagger(t)\, S\, U(t)
\end{align}
is independent of time, then the system relaxes to effective thermal equilibrium with the bath, so long as the bath itself is thermal.  We will examine whether such a scenario is possible when the system and bath are allowed to exchange particles via $H_{\rm SB}$.

\section{Effective thermalization and the grand canonical ensemble}
We will now consider whether and under what conditions the equilibration of a periodically driven system to an effective grand canonical distribution is possible.  As we wish to study cases where the total number of particles is conserved, while transport between the system and bath is allowed, we will assume the system-bath coupling to be of a similar form to that of Eq.~\eqref{grand canonical example}, namely
\begin{align}\label{HSB grand canonical}
H_{\rm SB}&=\sum_{\alpha,n} \(g_{\alpha}\, c^\dagger_{\alpha}\, b_{\alpha,n}+g^*_{\alpha}\, b^\dagger_{\alpha,n}\, c_{\alpha}\),
\end{align}
where $\alpha$ is a superindex that runs over all quantum numbers for the system (band index, momentum, etc.), and $n$ is a mode index for the bath.  All bath modes are assumed to couple equally to the system, although this requirement could be relaxed without altering our conclusion.  For concreteness, we take the system and bath to consist of fermions, although a derivation completely analogous to the one below can be carried out for bosons.

Let us now examine the conditions under which the system-bath coupling \eqref{HSB grand canonical} could remain time-independent in the basis defined by the time-dependent unitary transformation $U(t)$ that transforms $H_{\rm S}(t)$ into $H_{\rm eff}$.  For simplicity, we restrict ourselves to noninteracting Hamiltonians $H_{\rm S}(t)$.  In this case, the time-dependent unitary transformation $U(t)$ can be written in the form
\begin{align}
U(t)=\exp\[-\mathrm i\sum_{\alpha,\beta}h_{\alpha\beta}(t)\, c^\dagger_{\alpha}c_{\beta}\],
\end{align}
where $h_{\alpha\beta}(t)=h^*_{\beta\alpha}(t)$ defines a Hermitian matrix $h(t)$ of time-dependent coefficients.  
The transformed system-bath coupling is then
\begin{subequations}
\begin{align}
H_{\rm SB,\, eff}&\equiv U^\dagger(t)\, H^{\,}_{\rm SB}\, U(t) \nonumber \\
&=\sum_{\alpha,\beta,n} \(\[g_{\alpha}\, u^*_{\alpha\beta}(t)\, c^\dagger_{\beta}\]\, b_{\alpha,n}+ \text{H.c.}\),
\end{align}
where
\begin{align}
u_{\alpha\beta}(t)=\[e^{-\mathrm i\, h(t)}\]_{\alpha\beta}.
\end{align}
\end{subequations}
We would like to identify the conditions under which the effective system-bath coupling is independent of time.  To do this, it is convenient to make the Fourier expansions
\begin{subequations}
\begin{align}
u_{\alpha\beta}(t)=\sum_n e^{+\mathrm i\, n\Omega\, t}\, u^n_{\alpha\beta}\\
u^*_{\alpha\beta}(t)=\sum_n e^{-\mathrm i\, n\Omega\, t}\, u^{n\, *}_{\alpha\beta},
\end{align}
\end{subequations}
which reveal that $H_{\rm SB,\, eff}$ is independent of time when
\begin{align}\label{criterion}
g_\alpha^*\, u^n_{\alpha\beta}=0\indent \forall\, n\neq 0
\end{align}
for all $\alpha,\beta$.  This restrictive condition indicates that the driven system can only achieve effective equilibrium with the bath when the system-bath couplings are fine-tuned.  Furthermore, there is no guarantee that the above criterion can be satisfied for an arbitrary periodic driving protocol.  For example, the criterion \eqref{criterion} automatically fails for any driving protocol for which $u^0_{\alpha\beta}=0$.  Therefore, effetive thermalization for Floquet systems in the grand canonical ensemble also requires that the driving protocol induce a sufficiently simple transformation $U(t)$, which is generically not the case.

\section{Driven two-species model with dissipation}
We now illustrate the success and failure of effective thermalization for a simple class of models in which the transformation $U(t)$ is known exactly.  In particular, we demonstrate the possibility of finding a set of system-bath couplings that satisfiy the criterion \eqref{criterion} and show that the system effectively equilibrates with the bath for this choice of couplings.  We then show that a much larger set of other, perhaps more natural, choices of system-bath couplings leads to a failure of effective thermalization---the steady state displays non-thermal populations and a persistent current at long times.

\subsection{The model}
Let us introduce a toy model, which is nevertheless quite general, that constitutes the simplest possible scenario in which the driven system is allowed to exchange particles with the bath.  The system consists of two species of noninteracting fermions that are coupled by a driving field that excites transitions between the species:
\begin{subequations}\label{H_S}
\begin{align}
H_{\rm S}(t) &=H_1+H_2+H_{\rm D}(t)\\
H_{i} &= \sum_{\alpha,\bm k}E^{\, i}_{\alpha,\bm k}\ c^{\, i\, \dagger}_{\alpha,\bm k}\, c^{\, i}_{\alpha,\bm k}\label{H_i}\\
H_{\rm D}(t) &= \sum_{\alpha,\bm k}\(\Delta\,  e^{-\mathrm i \Omega t}\, c^{\, 1\, \dagger}_{\alpha,\bm k}\, c^{\, 2}_{\alpha,\bm k}+\text{H.c.}\),\label{H_D}
\end{align}
\end{subequations}
where $i=1,2$ labels the species, $\alpha = 1,\dots,N_{\rm b}$ labels the single-particle energy bands, and $\bm k$ labels the momentum in arbitrary dimensions.  The species index $i$ could label, for example, the fermion's spin, in which case a driving term of the form \eqref{H_D} arises from coupling the fermions to a linearly-polarized electric field.  (Thinking of the linearly-polarized field as a superposition of right- and left-circularly polarized fields, one finds that selection rules arising from conservation of angular momentum dictate that transitions from, say, spin-up to spin-down must be accompanied by an interaction with only one or the other chiral component of the field.)  The bath is modeled as two infinite reservoirs of free fermions,
\begin{align}
H_{\rm B} =\sum_{i=1,2} \sum_{\alpha,n,\bm k}\omega^{\, i}_{\alpha,n,\bm k}\ b^{\, i\, \dagger}_{\alpha,n,\bm k}\, b^{\, i}_{\alpha,n,\bm k}\label{H_B},
\end{align}
where $n=1,\dots,N_{\rm m}\gg N_{\rm b}$ labels the number of fermionic modes in the reservoir for each momentum $\bm k$ and band $\alpha$.  We assume this number to be much larger than the number of fermionic bands accessible to the system.  Each of the two reservoirs couples to one of the fermionic species in the system via
\begin{align}
H_{\rm SB}=\sum_{i=1,2}\, \sum_{\alpha,n,\bm k} g_{i}\( c^{\, i\, \dagger}_{\alpha,\bm k}\, b^{\, i}_{\alpha, n,\bm k}+\text{H.c.}\),
\end{align}
which allows for tunneling between the system and the reservoirs.  (We have made the system-bath couplings $g_i$ real for convenience.)

The system Hamiltonian defined in Eqs.~\eqref{H_S} falls into an exceptional class of periodically driven systems that admits an exact solution by a mapping to a rotating frame,
\begin{align}\label{rotating frame}
R(t)=\exp\[-\mathrm i\, \frac{\Omega}{2}\, t\(c^{\, 1\, \dagger}_{\alpha,\bm k}\, c^{\, 1}_{\alpha,\bm k}-c^{\, 2\, \dagger}_{\alpha,\bm k}\, c^{\, 2}_{\alpha,\bm k}\)\],
\end{align}
Which leads to
\begin{subequations}
\begin{align}
\tilde H_{\rm S} &= R^\dagger(t)\[H_{\rm S}(t)-\mathrm i\, \partial^{\,}_t\]R(t) \nonumber\\
&=\sum_{\alpha,\bm k}\[\sum_{i}\widetilde E^{\, i}_{\alpha,\bm k}\ c^{\, i\, \dagger}_{\alpha,\bm k}\, c^{\, i}_{\alpha,\bm k}\label{H_i}+\(\Delta\  c^{\, 1\, \dagger}_{\alpha,\bm k}\, c^{\, 2}_{\alpha,\bm k}+\text{H.c.}\)\],
\end{align}
where
\begin{align}
\widetilde E^{\, i}_{\alpha,\bm k}=E^{\, i}_{\alpha,\bm k}+(-1)^i\ \Omega/2.
\end{align}
\end{subequations}
The system Hamiltonian can then be re-diagonalized for each $\alpha$ and $\bm k$ by the unitary change of variables
\begin{subequations}\label{c tilde}
\begin{align}
\begin{pmatrix}
\tilde c^{\, 1}_{\alpha,\bm k}\\
\tilde c^{\, 2}_{\alpha,\bm k}
\end{pmatrix}
&=
W_{\alpha,\bm k}
 \begin{pmatrix}
c^{\, 1}_{\alpha,\bm k}\\
c^{\, 2}_{\alpha,\bm k},
\end{pmatrix}
\end{align}
in terms of which the system Hamiltonian becomes
\begin{align}\label{H S tilde}
\tilde H_{\rm S}&=\sum_{i=1,2}\sum_{\alpha,\bm k}\varepsilon^{\, i}_{\alpha,\bm k}\ \tilde c^{\, i\, \dagger}_{\alpha,\bm k}\, \tilde c^{\, i}_{\alpha,\bm k}.
\end{align}
\end{subequations}
\begin{subequations}
The rotating-frame eigenvalues $\varepsilon^i_{\alpha,\bm k}$ are given by
\begin{align}
\varepsilon^{\, i}_{\alpha,\bm k}&= \bar E_{\alpha,\bm k}+(-1)^i\ \delta E_{\alpha,\bm k},
\end{align}
where we have defined
\begin{align}
\bar E_{\alpha,\bm k}&=\frac{E^{\, 1}_{\alpha,\bm k}+E^{\, 2}_{\alpha,\bm k}}{2}\\
\delta E_{\alpha,\bm k}&=\sqrt{(\epsilon_{\alpha,\bm k}-\Omega/2)^2+|\Delta|^2}\\
\epsilon_{\alpha,\bm k}&=\frac{E^{\, 1}_{\alpha,\bm k}-E^{\, 2}_{\alpha,\bm k}}{2}
\end{align}
\end{subequations}
Under the transformation $R(t)$ defined in Eq.~\eqref{rotating frame}, the system-bath coupling acquires a time dependence:
\begin{align}\label{H SB tilde before basis change}
\tilde H_{\rm SB}(t)&=\sum_{\alpha,n,\bm k} \[g_1\(e^{+\mathrm{i}\Omega t/2}\, c^{\, 1\, \dagger}_{\alpha,\bm k}\, b^{\, 1}_{\alpha,n,\bm k}+\text{H.c.}\) \right.\nonumber \\
&\qquad\qquad \left.+g_2\(e^{-\mathrm{i}\Omega t/2}\, c^{\, 2\, \dagger}_{\alpha,\bm k}\, b^{\, 2}_{\alpha,n,\bm k}+\text{H.c.}\)\].
\end{align}
This time-dependent system-bath coupling can be handled without any assumptions about the magnitude of the driving frequency or the hybridization strength $|\Delta|$, as we will do in the next two sections.

\subsection{Effective thermalization when $g_2=0$}
Let us first treat the system-bath coupling in the manner outlined in the previous section, i.e.~by finding the time-periodic transformation $U(t)$ that renders the Hamiltonian $H_{\rm S}(t)$ time-independent.  Note that the Hamiltonian $\tilde H_{\rm S}$ in Eq.~\eqref{H S tilde} is already time-dependent, but that the transformation $R(t)$ defined in Eq.~\eqref{rotating frame} is not periodic with period $T=2\pi/\Omega$.  (It is, however, periodic with period $2T$.)  If instead we use
\begin{align}
U(t) = R(t)\, \exp\[+\mathrm i\, \frac{\Omega}{2}\, t\(c^{\, 1\, \dagger}_{\alpha,\bm k}\, c^{\, 1}_{\alpha,\bm k}+c^{\, 2\, \dagger}_{\alpha,\bm k}\, c^{\, 2}_{\alpha,\bm k}\)\],
\end{align}
 then we obtain
 \begin{subequations}
\begin{align}
H_{\rm eff}&=U^\dagger(t)\[H_{\rm S}(t)-\mathrm i\, \partial^{\,}_t\]U(t) \nonumber \\
&=\sum_{\alpha,\bm k}\[\sum_{i}\widetilde E^{\, \prime \, i}_{\alpha,\bm k}\ c^{\, i\, \dagger}_{\alpha,\bm k}\, c^{\, i}_{\alpha,\bm k}\label{H_i}+\(\Delta\  c^{\, 1\, \dagger}_{\alpha,\bm k}\, c^{\, 2}_{\alpha,\bm k}+\text{H.c.}\)\],
\end{align}
where now 
\begin{align}
\widetilde E^{\,\prime\, i}_{\alpha,\bm k}=E^{\, i}_{\alpha,\bm k}+\delta_{i,2}\,  \Omega.
\end{align}
\end{subequations}
$H_{\rm eff}$ is nothing but the Floquet effective Hamiltonian.  It is diagonal in the basis defined in Eq.~\eqref{c tilde}, and its eigenvalues $\varepsilon^{\, \prime\, i}_{\alpha,\bm k}$ (the Floquet quasienergies) are shifted with respect to those of $\tilde H_{\rm S}$ by $\Omega/2$:
\begin{align}
\varepsilon^{\, \prime\, i}_{\alpha,\bm k} &= \varepsilon^{\, i}_{\alpha,\bm k}+\frac{\Omega}{2}.
\end{align}
In this transformed reference frame, the system-bath coupling becomes
\begin{align}
H_{\rm SB,\,eff}(t)&=\sum_{\alpha,n,\bm k} \[g_1\(c^{\, 1\, \dagger}_{\alpha,\bm k}\, b^{\, 1}_{\alpha,n,\bm k}+\text{H.c.}\)\right.\nonumber\\&\qquad\qquad \left.+g_2\(e^{-\mathrm{i}\Omega t}\, c^{\, 2\, \dagger}_{\alpha,\bm k}\, b^{\, 2}_{\alpha,n,\bm k}+\text{H.c.}\)\].
\end{align}
We immediately see that $H_{\rm SB,\,eff}$ is time-independent along the line $g_2=0$ in the $g_1$-$g_2$ plane.  This is because, in the basis defined by the spinor $$\(c^1_{\alpha,\bm k}\ c^2_{\alpha,\bm k}\)^\textsf{T},$$ we have
\begin{align}
U(t)&=\begin{pmatrix}
1&0\\
0& e^{+\mathrm i\, \Omega\, t} 
\end{pmatrix}
=
\underbrace{\begin{pmatrix}
1&0\\
0&0
\end{pmatrix}}_{u^0}
+
e^{+\mathrm i\, \Omega\, t}
\underbrace{\begin{pmatrix}
0&0\\
0&1
\end{pmatrix}}_{u^1}.
\end{align}
Since there is only one Fourier component with $n\neq 0$, it suffices to choose the system-bath coupling vector $(g_1\ g_2)^\textsf{T}$ to belong to the left null space of $u^1$, which is just the line $g_2=0$.

\subsubsection{Derivation of the master equation}
To show that the system effectively thermalizes for $g_2=0$, let us first rewrite the system-bath coupling in the basis \eqref{c tilde} that diagonalizes $H_{\rm eff}$:
\begin{subequations}
\begin{align}
H_{\rm SB,\,eff}&=\sum_{i,\alpha,n,\bm k} \tilde g^{\, i}_{\alpha,\bm k}\(\tilde c^{\, i\, \dagger}_{\alpha,\bm k}\, b^{\, 1}_{\alpha,n,\bm k}+\text{H.c.}\),
\end{align}
where $i=1,2$ and
\begin{align}
\tilde g^1_{\alpha,\bm k}&=g_1\ \frac{\(\epsilon_{\alpha,\bm k}-\frac{\Omega}{2}\)-\delta E_{\alpha,\bm k}}{N^-_{\alpha,\bm k}}\\
\tilde g^2_{\alpha,\bm k}&=g_1\ \frac{\(\epsilon_{\alpha,\bm k}-\frac{\Omega}{2}\)+\delta E_{\alpha,\bm k}}{N^+_{\alpha,\bm k}},
\end{align}
with
\begin{align}
N^{ \pm}_{\alpha,\bm k} &= \sqrt{\[\(\epsilon_{\alpha,\bm k}-\Omega/2\)\pm\delta E_{\alpha,\bm k}\]^2+|\Delta|^2}.
\end{align}
\end{subequations}
We will now proceed by deriving the Born-Markov master equation for the evolution of the driven dissipative system.  

In the Born-Markov approximation, the total density matrix $\rho(t)$ is approximated as
\begin{subequations}
\begin{align}
\begin{split}
\rho(t) &\approx \rho_{\rm S}(t)\, \rho_B,\\
\rho_{\rm S}(t) &=\text{Tr}_{\rm B}\[\rho(t)\],\\
\rho_B &= \frac{e^{-\beta( H_{\rm B}-\mu\, N_{\rm B})}}{\text{Tr}[e^{-\beta( H_{\rm B}-\mu\, N_{\rm B})}]},
\end{split}
\end{align}
where $\text{Tr}_{\rm B}\[\ \cdot\ \]$ denotes the trace over bath degrees of freedom, and 
\begin{align}
N_{\rm B}=\sum_{\alpha,n,\bm k}b^{1\, \dagger}_{\alpha,n,\bm k}b^1_{\alpha,n,\bm k}
\end{align}
\end{subequations}
is the number operator for the bath.  Note that we have assumed that the bath is in thermal equilibrium at temperature $1/\beta$ and chemical equilibrium at chemical potential $\mu$ for all time.  The total density matrix solves the Liouville-von Neumann equation
\begin{align}
\dot\rho(t)&=-\mathrm i\, \[H(t),\rho(t)\],
\end{align}
where $H(t)$ is defined in Eq.~\eqref{H total}.  Transforming this equation using
\begin{align}
\tilde\rho_{\rm S}(t)&=U^\dagger(t)\, \rho_{\rm S}(t)\, U(t),
\end{align}
we obtain
\begin{align}\label{tilde rho dot}
\dot{\tilde \rho}(t)&=-\mathrm i\, [\tilde H(t),\tilde\rho(t)].
\end{align}
We next transform the left and right-hand sides to the interaction picture defined by
\begin{align}
\begin{split}
\mathcal O_{I}(t)&=U_{\rm S+B}^\dagger(t)\, \mathcal O\, U_{\rm S+B}(t)\\
U_{\mathrm{S+B}}(t) &= \mathcal T \exp\left\{-\mathrm i\int_0^t\mathrm d\pri t\, \[H_{\rm S}(\pri t)+H_{\rm B}\]\right\},
\end{split}
\end{align}
where the symbol $\mathcal T$ denotes time ordering.  Integrating both sides of Eq.~\eqref{tilde rho dot} with respect to time and working iteratively to leading order in the system-bath coupling, one finds the usual master equation
\begin{align}\label{reduced eom}
\dot{\tilde\rho}_{\mathrm S,I}(t)&=-\hspace{-.1cm}\int_0^\infty\hspace{-.15cm}\mathrm d\pri t\ \text{Tr}_{\rm B}\hspace{-.05cm}\left\{\[\tilde H_{\mathrm{SB},I}(t),\[\tilde H_{\mathrm{SB},I}(\pri t),\tilde\rho_I( t)\]\]\right\}.
\end{align}
The derivation of the above equation also entails the assumption that the correlation time of the bath is much smaller than any timescale associated with the system, including the driving period $T=2\pi/\Omega$.  This Markovian assumption allows one to neglect memory effects and extend the upper limit of the time integral to infinity.

The trace over the bath is carried out under the assumption that the bath degrees of freedom are well-described by the equilibrium grand canonical ensemble.  This implies, for instance, that
\begin{align}
\begin{split}
&\text{Tr}_{\rm B}\[b^{\, 1}_{\alpha,n,\bm k,I}(t)\, b^{\, 1}_{\beta,m,\pri{\bm k}\hspace{-.06cm},I}(\pri t)\, \rho_B\]\\
 &\qquad=\text{Tr}_{\rm B}\[b^{\, 1\, \dagger}_{\alpha,n,\bm k,I}(t)\, b^{\, 1\, \dagger}_{\beta,m,\pri{\bm k}\hspace{-.06cm},I}(\pri t)\, \rho_B\]=0,
\end{split}
\end{align}
and that
\begin{widetext}
\begin{subequations}
\begin{align}
\begin{split}
\text{Tr}_{\rm B}\[b_{\alpha,n,\bm k,I}^{\, 1\, \dagger}(t)\, b^{\, 1}_{\beta,m,\pri{\bm k}\hspace{-.06cm},I}(\pri t)\, \rho_{\rm B}\]&=\delta_{\alpha\beta}\,\delta_{mn}\,\delta_{\bm k\pri{\bm k}} \ n_F(\omega^{\, 1}_{\beta,m,\pri{\bm k}})\, e^{+\mathrm i\, \omega^{\, 1}_{\beta,m,\pri{\bm k}}(t-\pri t)}\\
\text{Tr}_{\rm B}\[b^{\, 1}_{\alpha,n,\bm k,I}(t)\, b^{\, 1\, \dagger}_{\beta,m,\pri{\bm k}\hspace{-.06cm},I}(\pri t)\, \rho_{\rm B}\]&=\delta_{\alpha\beta}\,\delta_{mn}\,\delta_{\bm k\pri{\bm k}} \ [1-n_F(\omega^{\, 1}_{\beta,m,\pri{\bm k}})]\, e^{-\mathrm i\, \omega^{\, 1}_{\beta,m,\pri{\bm k}}(t-\pri t)},
\end{split}
\end{align}
where
\begin{align}
n_{F}(E) = \frac{1}{e^{\beta(E-\mu)}+1}
\end{align}
is the Fermi-Dirac distribution.  The final quantity that characterizes the bath is the density of states
\begin{align}
\nu_{\alpha,\bm k}(E)&=\sum_{n}\delta(E-\omega^1_{\alpha,n,\bm k}),
\end{align}
which appears upon performing the time integration in Eq.~\eqref{reduced eom}.
\end{subequations}

Once the trace over the bath is completed and the operators are transformed back to the Schr\"odinger picture, we find that the Born-Markov master equation takes the form
\begin{subequations}
\begin{align}\label{lindblad master equation}
\dot{\tilde\rho}^{\,}_{\rm S}(t)&=-\mathrm i\, \[\tilde H^{\,}_{\rm S},\tilde \rho^{\,}_{\rm S}(t)\]+\sum_{\alpha, i,j,\bm k}\pi\, \(\tilde g^{\, i}_{\alpha,\bm k}\,\tilde g^{\, j}_{\alpha,\bm k}\) \, \nu^{\, j}_{\alpha,\bm k}\left\{n^{\, j}_{F,\alpha,\bm k}\ \mathcal D^{\, ij}_{\alpha,\bm k}\[\tilde c^{ \dagger}\]\, \tilde \rho^{\,}_{\rm S}(t)+\(1-n^{\, j}_{F,\alpha,\bm k}\)\, \mathcal D^{\, ij}_{\alpha,\bm k}\[\tilde c\]\, \tilde \rho^{\,}_{\rm S}(t)\right\},
\end{align}
where we have defined the shorthand
\begin{align}
\nu^{\, i}_{\alpha,\bm k}&=\nu_{\alpha,\bm k}(\varepsilon^{\, \prime\, i}_{\alpha,\bm k})\\
n^{\, i}_{F,\alpha,\bm k}&=n_F(\varepsilon^{\,\prime\, i}_{\alpha,\bm k}),
\end{align}
and the Lindbladian dissipators
\begin{align}
\mathcal D^{\, ij}_{\alpha,\bm k}\[\tilde c\]\, \tilde \rho^{\,}_{\rm S}(t)&=\tilde c^{\, i}_{\alpha,\bm k}\, \tilde \rho^{\,}_{\rm S}(t)\, \tilde c^{\, j\, \dagger}_{\alpha,\bm k}-\tilde c^{\, i\, \dagger}_{\alpha,\bm k}\, \tilde c^{\, j}_{\alpha,\bm k}\, \tilde \rho^{\,}_{\rm S}(t) + \text{H.c.}\\
\mathcal D^{\, ij}_{\alpha,\bm k}\[\tilde c^{ \dagger}\]\, \tilde \rho^{\,}_{\rm S}(t)&=\tilde c^{\, i\, \dagger}_{\alpha,\bm k}\, \tilde \rho^{\,}_{\rm S}(t)\, \tilde c^{\, j}_{\alpha,\bm k}-\tilde c^{\, i}_{\alpha,\bm k}\, \tilde c^{\, j\, \dagger}_{\alpha,\bm k}\, \tilde \rho^{\,}_{\rm S}(t) + \text{H.c.}
\end{align}
\end{subequations}
The first term in Eq.~\eqref{lindblad master equation} describes the unitary evolution of the reduced density matrix due to the dynamics of the system alone, while the second term arises due to the non-unitary evolution obtained as a result of the trace over the bath.  It is a well-known result of quantum statistical mechanics that, despite their non-unitary nature, master equations of the form \eqref{lindblad master equation} preserve the trace of the density matrix.~\cite{lindblad}

\subsubsection{Steady-state populations and coherences}
Eq.~\eqref{lindblad master equation} can be used to calculate the time evolution of the expectation value of any Schr\"odinger-picture observable $\mathcal O$ as follows.  First note that
\begin{subequations}
\begin{align}
\der{}{t}\expect{\mathcal O}_{\tilde\rho^{\,}_{\rm S}}\equiv\der{}{t}\,\text{Tr}\[\mathcal O\, \tilde\rho^{\,}_{\rm S}(t)\]=\text{Tr}\[\mathcal O\, \dot{\tilde\rho}^{\,}_{\rm S}(t)\].
\end{align}
The unitary contribution to the time evolution of $\expect{\mathcal O}$ is obtained as
\begin{align}
\text{Tr}\left\{\mathcal O\, \[\tilde H^{\,}_{\rm S},\tilde\rho^{\,}_{\rm S}(t)\]\right\}=\expect{\[\mathcal O,\tilde H^{\,}_{\rm S}\]}_{\tilde \rho^{\,}_{\rm S}},
\end{align}
where we used the cyclic property of the trace.  The dissipative contribution comes from further application of the cyclic property to
\begin{align}
\begin{split}
\text{Tr}\left\{\mathcal O\  \mathcal D^{\, ij}_{\alpha,\bm k}\[\tilde c^{ \dagger}\]\, \tilde \rho^{\,}_{\rm S}(t)\right\}&= \expect{\[\tilde c^{\, i}_{\alpha, \bm k}, \mathcal O\]\tilde c^{\, j\, \dagger}_{\alpha,\bm k}}_{\tilde \rho^{\,}_{\rm S}}+\expect{\tilde c^{\, j}_{\alpha, \bm k}\, \[\mathcal O,\tilde c^{\, i\, \dagger}_{\alpha,\bm k}\]}_{\tilde \rho^{\,}_{\rm S}}\\
\text{Tr}\left\{\mathcal O\  \mathcal D^{\, ij}_{\alpha,\bm k}\[\tilde c\]\, \tilde \rho^{\,}_{\rm S}(t)\right\}&= \expect{\[\tilde c^{\, i\, \dagger}_{\alpha, \bm k}, \mathcal O\]\tilde c^{\, j}_{\alpha,\bm k}}_{\tilde \rho^{\,}_{\rm S}}+\expect{\tilde c^{\, j\, \dagger}_{\alpha, \bm k}\, \[\mathcal O,\tilde c^{\, i}_{\alpha,\bm k}\]}_{\tilde \rho^{\,}_{\rm S}}.
\end{split}
\end{align}
\end{subequations}

To determine the equations of motion for the populations and coherences, we apply the above formulas for the choice $\mathcal O=c^{\, p\, \dagger}_{\alpha,\bm k}\, c^{\, q}_{\alpha,\bm k}$ and obtain
\begin{subequations}\label{EOM populations g2=0}
\begin{align}
\dot{\tilde n}^{11}_{\alpha,\bm k}&= -2\pi\, (\tilde g^{1}_{\alpha,\bm k})^2\, \nu^{1}_{\alpha,\bm k}\, \(\tilde n^{11}_{\alpha,\bm k}-n^{1}_{F,\alpha,\bm k}\)-\pi\, \nu^{2}_{\alpha,\bm k}\, \tilde g^{1}_{\alpha,\bm k}\, \tilde g^{2}_{\alpha,\bm k}\, \(\tilde n^{12}_{\alpha,\bm k}+\tilde n^{21}_{\alpha,\bm k}\) \\
\dot{\tilde n}^{22}_{\alpha,\bm k}&= -2\pi\, (\tilde g^{2}_{\alpha,\bm k})^2\, \nu^{2}_{\alpha,\bm k}\, \(\tilde n^{22}_{\alpha,\bm k}-n^{2}_{F,\alpha,\bm k}\)-\pi\, \nu^{1}_{\alpha,\bm k}\, \tilde g^{2}_{\alpha,\bm k}\, \tilde g^{1}_{\alpha,\bm k} \(\tilde n^{12}_{\alpha,\bm k}+\tilde n^{21}_{\alpha,\bm k}\)\\
\begin{split}
\dot{\tilde n}^{12}_{\alpha,\bm k}&=+\mathrm i\, \(\varepsilon^{\prime\, 1}_{\alpha,\bm k}-\varepsilon^{\prime\, 2}_{\alpha,\bm k}\)\, \tilde n^{12}_{\alpha,\bm k}-\pi\left[(\tilde g^{1}_{\alpha,\bm k})^2\, \nu^{1}_{\alpha,\bm k}+(\tilde g^{2}_{\alpha,\bm k})^2\, \nu^{2}_{\alpha,\bm k} \right]\, \tilde n^{12}_{\alpha,\bm k} \\
&\indent\indent -\pi\, \tilde g^{2}_{\alpha,\bm k}\, \tilde g^{1}_{\alpha,\bm k}\, \nu^{1}_{\alpha,\bm k}\, (\tilde n^{11}_{\alpha,\bm k}-n^{1}_{F,\alpha,\bm k})-\pi\, \tilde g^{1 }_{\alpha,\bm k}\, \tilde g^{2 }_{\alpha,\bm k}\, \nu^{2 }_{\alpha,\bm k}\, (\tilde n^{22}_{\alpha,\bm k}-n^{2}_{F,\alpha,\bm k})
\end{split}
\\
\dot{\tilde n}^{12}_{\alpha,\bm k}&=(\dot{\tilde n}^{21}_{\alpha,\bm k})^*,
\end{align}
\end{subequations}
\end{widetext}
where we have defined
\begin{align}
\tilde n^{\, ij}_{\alpha,\bm k}&=\expect{\tilde c^{\, i\, \dagger}_{\alpha,\bm k}\, \tilde c^{\, j}_{\alpha,\bm k}}_{\tilde\rho_{\rm S}}.
\end{align}
The steady-state values of the populations and coherences correspond to fixed-points of the equations of motion \eqref{EOM populations g2=0}.  They are obtained by setting $\dot{\tilde n}^{\, pq}_{\alpha,\bm k}=0$ for all $p$ and $q$ and solving the resulting linear equations.  We find that the equations of motion~\eqref{EOM populations g2=0} have a unique, stable fixed-point solution where the system reaches effective thermal and chemical equilibrium with the reservoir:
\begin{align}\label{thermal distribution}
\begin{split}
\tilde n^{\, 11}_{\alpha,\bm k}&=n_{F}(\varepsilon^{\prime \, 1}_{\alpha,\bm k})\\
\tilde n^{\, 22}_{\alpha,\bm k}&=n_{F}(\varepsilon^{\prime \, 2}_{\alpha,\bm k})\\
\tilde n^{\, 12}_{\alpha,\bm k}&=0.
\end{split}
\end{align}
Observe that this effective equilibrium is characterized by the same temperature and chemical potential as the reservoir, as it would be in a static system.

\subsection{Failure of effective thermalization when $g_1,g_2\neq 0$}
The long-time steady state of the system can still be found for arbitrary $g_1$ and $g_2$, although we will see that in this case the system does not equilibrate to an effective steady-state distribution with the same temperature and chemical potential as the bath.  Rewriting Eq.~\eqref{H SB tilde before basis change} in terms of the operators $\tilde c^{\, i}_{\alpha,\bm k}$ defined in Eq.~\eqref{c tilde} above, we find that the system-bath interaction takes the form
\begin{align}\label{H SB tilde}
\tilde H_{\rm SB}(t)=\sum_{i,j}\, \sum_{\alpha,n,\bm k} \(\tilde g^{ij}_{\alpha,\bm k}\ \tilde c^{\, i\, \dagger}_{\alpha,\bm k}\, \tilde b^{\, j}_{\alpha,n,\bm k}(t)+\text{H.c.}\),
\end{align}
where the system-bath coupling matrix is given by
\begin{align}
\tilde g_{\alpha,\bm k}^{ij} = W_{\alpha,\bm k}^{ij}\, g_j
\end{align}
(no implicit sum on repeated indices), and where we have defined the explicitly time-dependent operators
\begin{align}\label{b tilde definition}
\tilde b^{\, j}_{\alpha,n,\bm k}(t)&=\exp\[-\mathrm i\, (-1)^j\, \Omega\, t/2\]\, b^{\, j}_{\alpha,n,\bm k}.
\end{align}

\subsubsection{Derivation of the master equation}
The derivation of the Born-Markov master equation for the time evolution of the reduced density matrix proceeds almost identically to the one sketched in the previous section.  We therefore point out here some of the more important changes brought about by the time-dependent system-bath couplings in Eq.~\eqref{H SB tilde}.

First, since the system is effectively coupled to two reservoirs (one for each species), the additional assumption that both reservoirs are uncorrelated and independently thermalized on time scales associated with the system's dynamics implies that
\begin{align}
\begin{split}
&\text{Tr}_{\rm B}\[\tilde b^{\, i}_{\alpha,n,\bm k,I}(t)\, \tilde b^{\, j}_{\beta,m,\pri{\bm k}\hspace{-.06cm},I}(\pri t)\, \rho_B\]\\
&\qquad=\text{Tr}_{\rm B}\[\tilde b^{\, i\, \dagger}_{\alpha,n,\bm k,I}(t)\, \tilde b^{\, j\, \dagger}_{\beta,m,\pri{\bm k}\hspace{-.06cm},I}(\pri t)\, \rho_B\]=0.
\end{split}
\end{align}
Second, the bath correlators are now given by
\begin{widetext}
\begin{subequations}
\begin{align}
\expect{\tilde{b}_{\alpha,n,\bm k,I}^{\, j\, \dagger}(t)\, \tilde{b}^{\, l}_{\beta,m,\pri{\bm k},I}(\pri t)}_{\rho_{\rm B}}&=\delta_{\alpha\beta}\delta_{mn}\delta_{\bm k\pri{\bm k}}\delta_{jl}\ n_F(\omega^{\, l}_{\beta,m,\pri{\bm k}})\, e^{+\mathrm i\[\omega^{\, l}_{\beta,m,\pri{\bm k}}+(-1)^l\, \frac{\Omega}{2}\](t-\pri t)}\\
\expect{\tilde{b}^{\, j}_{\alpha,n,\bm k,I}(t)\, \tilde{b}^{\, l\, \dagger}_{\beta,m,\pri{\bm k},I}(\pri t)}_{\rho_{\rm B}}&=\delta_{\alpha\beta}\delta_{mn}\delta_{\bm k\pri{\bm k}}\delta_{jl}\ [1-n_F(\omega^{\, l}_{\beta,m,\pri{\bm k}})]\, e^{-\mathrm i\[\omega^{\, l}_{\beta,m,\pri{\bm k}}+(-1)^l\, \frac{\Omega}{2}\](t-\pri t)}.
\end{align}
\end{subequations}
Of crucial importance above is the species-dependent shift of the bath energies by $\pm\Omega/2$, which appears in the exponentials but is not present in the Fermi-Dirac distributions.  These energy shifts appear as a result of the explicit time dependence of the bath operators $\tilde b^{\, j}_{\alpha,n,\bm k}(t)$ defined in Eq.~\eqref{b tilde definition}.  Consequently, the rotating-frame eigenvalues that appear in the bath density of states and Fermi-Dirac distributions upon tracing out the bath are also shifted by $\pm\Omega/2$.  These shifts have a profound effect on the equations of motion for the populations and coherences, which we now discuss.

\subsubsection{Steady-state populations and coherences}
The equations of motion for the populations and coherences are obtained by direct analogy with the analysis of the previous section.  We find
\begin{subequations}\label{EOM populations}
\begin{align}
\dot{\tilde n}^{11}_{\alpha,\bm k}&= -2\pi\sum_{l=1,2}|\tilde g^{1l}_{\alpha,\bm k}|^2\, \nu^{1l}_{\alpha,\bm k}\, \(\tilde n^{11}_{\alpha,\bm k}-n^{1l}_{F,\alpha,\bm k}\)-\pi\sum_{l=1,2}  \nu^{2 l}_{\alpha,\bm k}\, \(\tilde g^{1l}_{\alpha,\bm k}\, \tilde g^{2 l\, *}_{\alpha,\bm k}\, \tilde n^{12}_{\alpha,\bm k}+\tilde g^{1l\, *}_{\alpha,\bm k}\, \tilde g^{2 l}_{\alpha,\bm k}\,\tilde n^{21}_{\alpha,\bm k}\)\\
\dot{\tilde n}^{22}_{\alpha,\bm k}&= -2\pi\sum_{l=1,2}|\tilde g^{2l}_{\alpha,\bm k}|^2\, \nu^{2l}_{\alpha,\bm k}\, \(\tilde n^{22}_{\alpha,\bm k}-n^{2l}_{F,\alpha,\bm k}\)-\pi\sum_{l=1,2}  \nu^{1 l}_{\alpha,\bm k}\, \(\tilde g^{2l\, *}_{\alpha,\bm k}\, \tilde g^{1 l}_{\alpha,\bm k}\, \tilde n^{12}_{\alpha,\bm k}+\tilde g^{2l}_{\alpha,\bm k}\, \tilde g^{1 l\, *}_{\alpha,\bm k}\,\tilde n^{21}_{\alpha,\bm k}\) \\
\begin{split}
\dot{\tilde n}^{12}_{\alpha,\bm k}&=+\mathrm i\, \(\varepsilon^{1}_{\alpha,\bm k}-\varepsilon^2_{\alpha,\bm k}\)\, \tilde n^{12}_{\alpha,\bm k}-\pi\sum_{l=1,2} \(|\tilde g^{1l}_{\alpha,\bm k}|^2\, \nu^{1l}_{\alpha,\bm k}+|\tilde g^{2l}_{\alpha,\bm k}|^2\, \nu^{2l}_{\alpha,\bm k}\)\, \tilde n^{12}_{\alpha,\bm k} \\
&\indent\indent -\pi\sum_{l=1,2}\tilde g^{2l}_{\alpha,\bm k}\, \tilde g^{1 l\, *}_{\alpha,\bm k}\, \nu^{1 l}_{\alpha,\bm k}\, (\tilde n^{11}_{\alpha,\bm k}-n^{1l}_{F,\alpha,\bm k})-\pi\sum_{l=1,2} \tilde g^{1 l\, *}_{\alpha,\bm k}\, \tilde g^{2 l}_{\alpha,\bm k}\, \nu^{2 l}_{\alpha,\bm k}\, (\tilde n^{22}_{\alpha,\bm k}-n^{2l}_{F,\alpha,\bm k}) 
\end{split}
\\
\dot{\tilde n}^{12}_{\alpha,\bm k}&=(\dot{\tilde n}^{21}_{\alpha,\bm k})^*,
\end{align}
\end{subequations}
where we have defined the quantities
\begin{subequations}
\begin{align}
n^{\, ij}_{F,\alpha,\bm k} &= n_F\(\varepsilon^{\, i}_{\alpha,\bm k}-(-1)^j\, \Omega/2\)\\
\nu^{\, ij}_{\alpha,\bm k}&= \nu\(\varepsilon^{\, i}_{\alpha,\bm k}-(-1)^j\, \Omega/2\).
\end{align}
\end{subequations}

Eqs.~\eqref{EOM populations}, being linear in the $\tilde n^{\, ij}_{\alpha,\bm k}$, can be solved exactly for each $\alpha$ and $\bm k$.  However, the exact solution is a complicated function of the bare energy spectrum $E^{\, i}_{\alpha,\bm k}$, the hybridization strength $|\Delta|$, and the driving frequency $\Omega$.  Progress can be made by employing the following relatively mild assumptions about the bath:
\begin{enumerate}
\item The two baths are identical, i.e.~$\omega^1_{\alpha,n,\bm k}=\omega^2_{\alpha,n,\bm k}$, and the bare system-bath couplings are also identical, i.e.~$g_1=g_2=g$.
\item The bath spectra $\omega^{\, i}_{\alpha,n,\bm k}$ are unbounded.
\item The bath spectra $\omega^{\, i}_{\alpha,n,\bm k}$ are sufficiently dense with spacing $\Delta E$ that we can make the replacement $\sum_{n}\to\int (\mathrm dE)/\Delta E$.
\end{enumerate}
These assumptions reveal relations between the functions $\nu^{\, ij}_{\alpha,\bm k}$, namely
\begin{align}
\nu^{\, i1}_{\alpha,\bm k} &= \nu^{\, i2}_{\alpha,\bm k}.
\end{align}
Combining this relation with the assumption that $g_1=g_2$, one finds that
\begin{align}
\sum_{l=1,2}  \nu^{2 l}_{\alpha,\bm k}\, \tilde g^{1l}_{\alpha,\bm k}\, \tilde g^{2 l\, *}_{\alpha,\bm k} = \sum_{l=1,2}  \nu^{1 l}_{\alpha,\bm k}\, \tilde g^{2l\, *}_{\alpha,\bm k}\, \tilde g^{1 l}_{\alpha,\bm k} = 0.
\end{align}
This is an immense simplification, as it decouples Eqs.~\eqref{EOM populations}:
\begin{subequations}\label{EOM populations decoupled}
\begin{align}
\dot{\tilde n}^{11}_{\alpha,\bm k}&= -2\pi\sum_{l=1,2}|\tilde g^{1l}_{\alpha,\bm k}|^2\, \nu^{1l}_{\alpha,\bm k}\, \(\tilde n^{11}_{\alpha,\bm k}-n^{1l}_{F,\alpha,\bm k}\)\\
\dot{\tilde n}^{22}_{\alpha,\bm k}&= -2\pi\sum_{l=1,2}|\tilde g^{2l}_{\alpha,\bm k}|^2\, \nu^{2l}_{\alpha,\bm k}\, \(\tilde n^{22}_{\alpha,\bm k}-n^{2l}_{F,\alpha,\bm k}\)\\
\begin{split}
\dot{\tilde n}^{12}_{\alpha,\bm k}&=+\mathrm i\, \(\varepsilon^{1}_{\alpha,\bm k}-\varepsilon^2_{\alpha,\bm k}\)\, \tilde n^{12}_{\alpha,\bm k}-\pi\sum_{l=1,2} \(|\tilde g^{1l}_{\alpha,\bm k}|^2\, \nu^{1l}_{\alpha,\bm k}+|\tilde g^{2l}_{\alpha,\bm k}|^2\, \nu^{2l}_{\alpha,\bm k}\)\, \tilde n^{12}_{\alpha,\bm k}\\ 
&\hspace{1.5cm}+\pi\sum_{l=1,2}\tilde g^{2l}_{\alpha,\bm k}\, \tilde g^{1 l\, *}_{\alpha,\bm k}\, \nu^{1 l}_{\alpha,\bm k}\, n^{1l}_{F,\alpha,\bm k}+\pi\sum_{l=1,2} \tilde g^{1 l\, *}_{\alpha,\bm k}\, \tilde g^{2 l}_{\alpha,\bm k}\, \nu^{2 l}_{\alpha,\bm k}\, n^{2l}_{F,\alpha,\bm k}
\end{split}
\\
\dot{\tilde n}^{12}_{\alpha,\bm k}&=(\dot{\tilde n}^{21}_{\alpha,\bm k})^*,
\end{align}
\end{subequations}
The steady-state solution to these equations arises from setting the left-hand side to zero.  The steady-state populations of the Floquet states are found to be
\begin{subequations}\label{steady-state solution}
\begin{align}
\tilde n^{11}_{\alpha,\bm k}&=\frac{\sum_{l=1,2}|\tilde g^{1l}_{\alpha,\bm k}|^2\, \nu^{1l}_{\alpha,\bm k}\ n^{1l}_{F,\alpha,\bm k}}{\sum_{l=1,2}|\tilde g^{1l}_{\alpha,\bm k}|^2\, \nu^{1l}_{\alpha,\bm k}} \label{population 1}\\
\tilde n^{22}_{\alpha,\bm k}&=\frac{\sum_{l=1,2}|\tilde g^{2l}_{\alpha,\bm k}|^2\, \nu^{2l}_{\alpha,\bm k}\ n^{2l}_{F,\alpha,\bm k}}{\sum_{l=1,2}|\tilde g^{2l}_{\alpha,\bm k}|^2\, \nu^{2l}_{\alpha,\bm k}}, \label{population 2}
\end{align}
while the coherences are found to be
\begin{align}\label{12 coherence}
\tilde n^{12}_{\alpha,\bm k}&=\frac{-\pi\sum_{l=1,2}\(\tilde g^{2l}_{\alpha,\bm k}\, \tilde g^{1 l\, *}_{\alpha,\bm k}\, \nu^{1 l}_{\alpha,\bm k}\ n^{1l}_{F,\alpha,\bm k}+\tilde g^{1 l\, *}_{\alpha,\bm k}\, \tilde g^{2 l}_{\alpha,\bm k}\, \nu^{2 l}_{\alpha,\bm k}\ n^{2l}_{F,\alpha,\bm k}\)}{\mathrm i\, (\varepsilon^1_{\alpha,\bm k}-\varepsilon^2_{\alpha,\bm k})-\pi\sum_{l=1,2} \(|\tilde g^{1l}_{\alpha,\bm k}|^2\, \nu^{1l}_{\alpha,\bm k}+|\tilde g^{2l}_{\alpha,\bm k}|^2\, \nu^{2l}_{\alpha,\bm k}\)}.
\end{align}
\end{subequations}
\end{widetext}
From the above solution, we see that, for arbitrary driving parameters $|\Delta|$ and $\Omega$, the system relaxes to a nonthermal distribution with nonvanishing coherences $\tilde n^{12}_{\alpha,\bm k}=\langle \tilde c^{\, 1\, \dagger}_{\alpha,\bm k}\, \tilde c^{\, 2}_{\alpha,\bm k}\rangle$.  Below, we provide intuitive interpretations for the above expressions.

\subsubsection{Nonthermal populations}

The expressions for the steady-state populations in Eqs.~\eqref{population 1} and \eqref{population 2} can be understood intuitively by taking the limit $|\Omega| \gg E^{\, i}_{\alpha,\bm k}, \Delta$ for all $i,\alpha,$ and $\bm k$.  To order $1/|\Omega|$, the steady-state populations are (see Appendix)
\begin{align}\label{simplest nonthermal populations}
\begin{split}
\tilde n^{11}_{\alpha,\bm k}&=n_{F}\(\varepsilon^{1}_{\alpha,\bm k}+\frac{|\Omega|}{2}\)\\
\tilde n^{22}_{\alpha,\bm k}&=n_{F}\(\varepsilon^{2}_{\alpha,\bm k}-\frac{|\Omega|}{2}\).
\end{split}
\end{align}
Evidently, only the $l=1$ ($l=2$) terms in Eq.~\eqref{population 1} survive in the limit $\Omega\to +\infty$ ($\Omega\to -\infty$), while only the $l=2$ ($l=1$) terms contribute to Eq.~\eqref{population 2}.~\cite{footnote:cuetara}  

In this limit, we see that the system relaxes to a steady-state distribution similar to the thermal distribution of Eq.~\eqref{thermal distribution}, except that the two species of fermions acquire an effective relative bias of magnitude $|\Omega|$ with respect to one another.  Thus, even in the limit of large driving frequency, the system remains out of equilibrium, since the two species equilibrate with different effective chemical potentials.  As the driving frequency is decreased, the populations become increasingly nonthermal---at order $1/|\Omega|^2$, for example, the $l=2$ ($l=1$) terms in Eq.~\eqref{population 1} [and, likewise, the $l=1$ ($l=2$) terms in Eq.~\eqref{population 2}] provide a small but finite contribution.  At arbitrary values of $\Omega$, the final steady-state distributions are weighted sums of the form
\begin{align}
\tilde n^{11}_{\alpha,\bm k} &= \frac{A_{\alpha,\bm k}\ n_F(\varepsilon^{1}_{\alpha,\bm k}+\frac{\Omega}{2})+B_{\alpha,\bm k}\ n_F(\varepsilon^{1}_{\alpha,\bm k}-\frac{\Omega}{2})}{A_{\alpha,\bm k}+B_{\alpha,\bm k}}\\
\tilde n^{22}_{\alpha,\bm k} &= \frac{\pri A_{\alpha,\bm k}\ n_F(\varepsilon^{2}_{\alpha,\bm k}-\frac{\Omega}{2})+\pri B_{\alpha,\bm k}\ n_F(\varepsilon^{2}_{\alpha,\bm k}+\frac{\Omega}{2})}{\pri A_{\alpha,\bm k}+\pri B_{\alpha,\bm k}},\nonumber
\end{align}
where the weights $A_{\alpha,\bm k},B_{\alpha,\bm k},\pri A_{\alpha,\bm k},$ and $\pri B_{\alpha,\bm k}$ depend on $|\Delta|$ and $\Omega$ and are comparable in size for $|\Omega|\sim E^{i}_{\alpha,\bm k},|\Delta|$.  (For explicit expressions, see the Appendix.)

\subsubsection{Persistent flavor current at long times}
The nonvanishing steady-state expectation value $\tilde n^{12}_{\alpha,\bm k}$ indicates the presence of a flavor current that persists at long times in the absence of an external bias.  The flavor current from species 1 to species 2 is defined as
\begin{align}\label{general current expression}
\begin{split}
J_{1\to 2} &= \sum_{\alpha,\bm k}\(\Delta \expect{c^{2\, \dagger}_{\alpha,\bm k}\, c^{1}_{\alpha,\bm k}} - \Delta^* \expect{c^{1\, \dagger}_{\alpha,\bm k}\, c^{2}_{\alpha,\bm k}}\)\\
&=\sum_{\alpha,\bm k}2\, \text{Im}\[\Delta\, \expect{c^{2\, \dagger}_{\alpha,\bm k}\, c^{1}_{\alpha,\bm k}}\]\\
&=\sum_{\alpha,\bm k}2\, |\Delta|\, \text{Im}\(\tilde n^{12}_{\alpha,\bm k}\)
\end{split}
\end{align}
In the case where $\nu^{\, ij}_{\alpha,\bm k}\equiv\nu=\text{const.}$ for all $i,j,\alpha,\bm k$, and where $g_1=g_2=g$, we have that (see Appendix)
\begin{align}
\begin{split}
\tilde g^{12}_{\alpha,\bm k}\,\tilde g^{22\, *}_{\alpha,\bm k}\, \nu^{22}_{\alpha,\bm k}=-\tilde g^{21\, *}_{\alpha,\bm k}\,\tilde g^{11}_{\alpha,\bm k}\, \nu^{11}_{\alpha,\bm k} &= g^2\, \nu\, \frac{|\Delta|}{2\, \delta E_{\alpha,\bm k}}\\
\sum_{l=1,2} \(|\tilde g^{1l}_{\alpha,\bm k}|^2\, \nu^{1l}_{\alpha,\bm k}+|\tilde g^{2l}_{\alpha,\bm k}|^2\, \nu^{2l}_{\alpha,\bm k}\) &= 2\, g^2\, \nu,
\end{split}
\end{align}
so that the current is given by
\begin{align}\label{J1->2}
\begin{split}
J_{1\to 2}&=-\pi\, g^2\, \nu\, |\Delta|^2\\
&\indent\times  \sum_{\alpha,\bm k}\, \frac{n^{22}_{F,\alpha,\bm k}-n^{21}_{F,\alpha,\bm k}-n^{11}_{F,\alpha,\bm k}+n^{12}_{F,\alpha,\bm k}}{(\delta E_{\alpha,\bm k})^2+(2\pi\, g^2\, \nu)^2},
\end{split}
\end{align}
in units such that the elementary charge $e=1$.  The presence of this persistent current at long times highlights the out-of-equilibrium nature of the steady state.  Due to the absence of an external bias, this current arises entirely due to the driving and to the fact that the coupling to the reservoirs allows for net particle transport.

In the next section, we will analyze the above expression for the current in the context of two physical examples, where the flavor current can be interpreted as a charge current in one case, and a valley current in the other.

\subsection{Physical examples}

To make the results of the previous section more concrete, we now study two physical examples of the model defined in Eq.~\eqref{H_S}.  A particular realization of this model amounts to making a choice of single-particle Hilbert space (i.e.~of the set of quantum numbers indexed by $\alpha$), and a choice of bare dispersion $E^{\, i}_{\alpha,\bm k}$.  For simplicity, we will restrict ourselves to the case where the reservoir has a constant density of states.  However, we note that the general expression for the current, given in Eq.~\eqref{general current expression} depends implicitly on the system-bath couplings and on the bath density of states via $\tilde n^{12}_{\alpha,\bm k}$ [c.f.~Eqs.~\eqref{steady-state solution}].  As such, different choices of bath densities of states will alter the findings below.  A detailed discussion of this dependence of the nonthermal occupations and persistent current on the bath density of states is a worthwhile undertaking, but is beyond the scope of this paper.

\begin{figure}[t]
(a)\includegraphics[width=.475\textwidth]{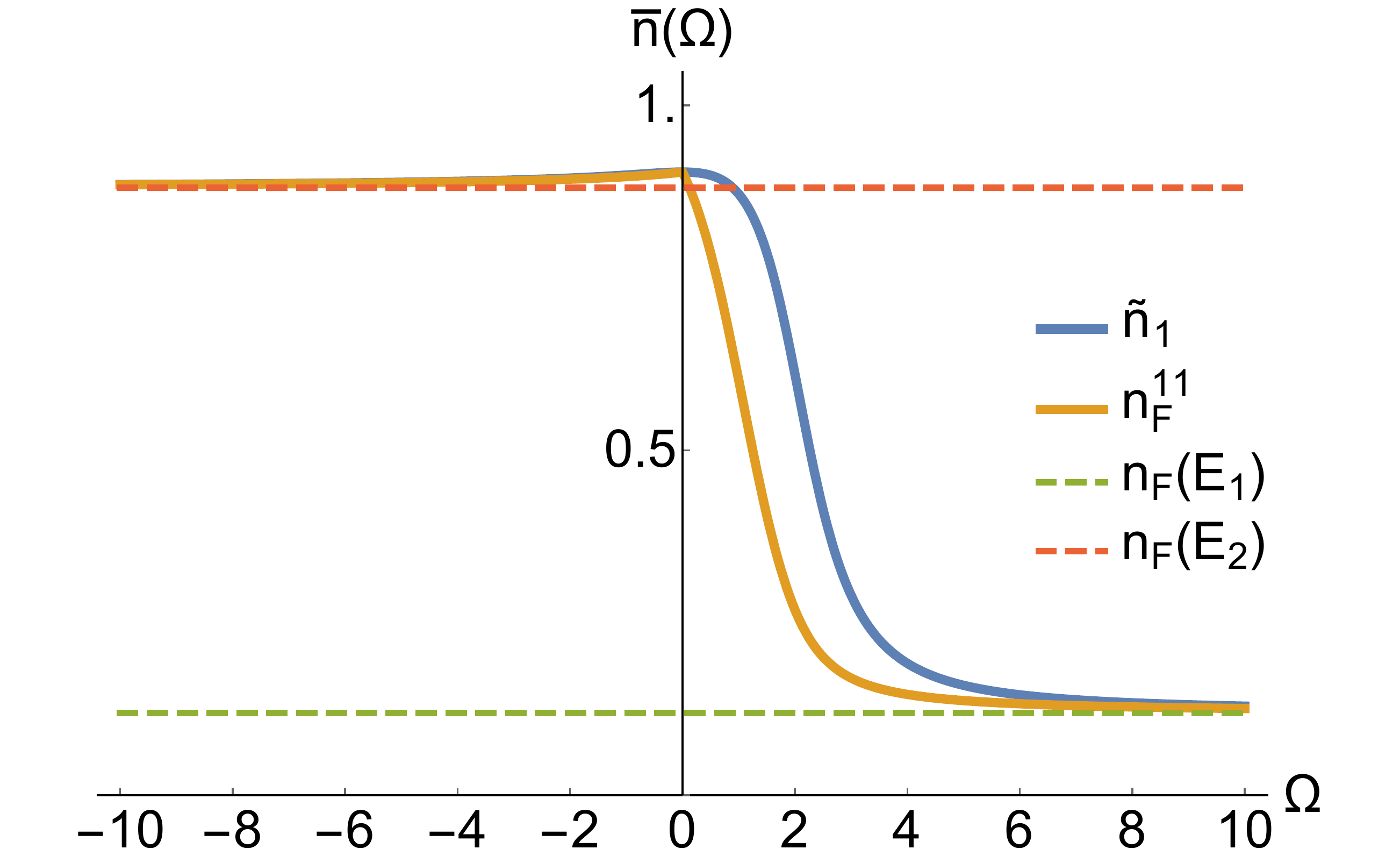}
\\
(b)\includegraphics[width=.475\textwidth]{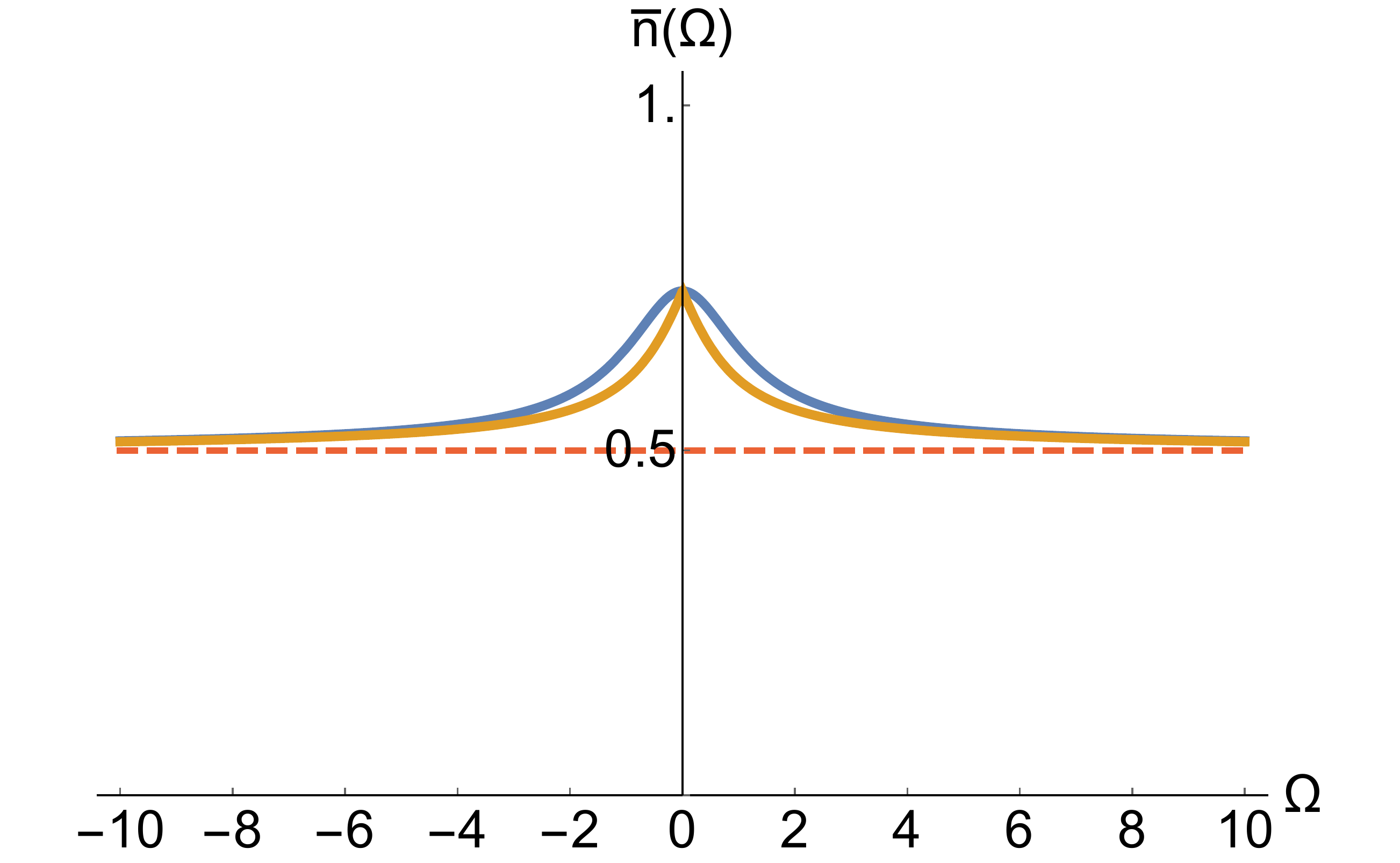}
\\
(c)\includegraphics[width=.475\textwidth]{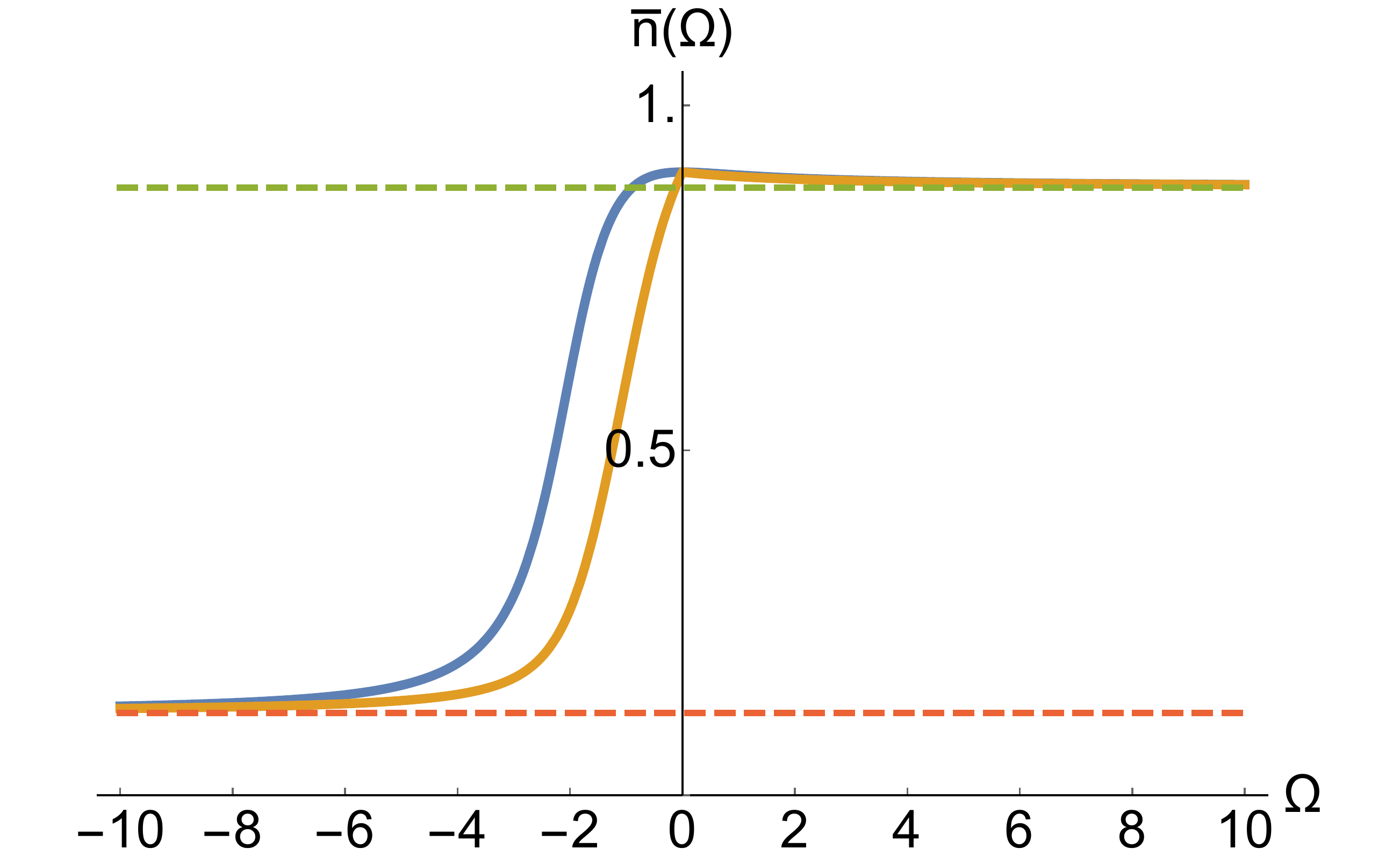}
\caption{(Color online) Average occupations $\bar n(\Omega)$ as a function of frequency in the double-dot example for detunings (a) $\epsilon=+1$, (b) $\epsilon=0$, and (c) $\epsilon=-1$.  All energies are measured in units of the rotating-frame gap $2|\Delta|$, and we take the inverse temperature $\beta=2$.  The asymptotic distributions given in Eq.~\eqref{simplest nonthermal populations}, as well as the equilibrium values of the occupations, are shown for reference.  
\label{fig:dot_pops}
        }
\end{figure}

\subsubsection{Driven double quantum dot}\label{double dot example}

Let us first consider perhaps the simplest example, namely that of a driven double quantum dot coupled to two spinless fermionic reservoirs.  In this example, the species label $i=1,2$ distinguishes two single-level quantum dots (see Figure~\ref{fig:double_dot}), each of which is coupled to its own reservoir at inverse temperature $\beta_{1,2}$ and chemical potential $\mu_{1,2}$.  (For the purposes of our discussion we will set $\beta_1=\beta_2=\beta$ and $\mu_1=\mu_2=\mu$, in order to focus on out-of-equilibrium effects due to the driving alone.)  We
suppress the labels $\alpha$ and $\bm k$ and define $E^{\, i}_{\alpha,\bm k}\equiv E_i$ to be the potential at which dot $i$ is held.  The eigenvalues of the Hamiltonian in the rotating frame (which, up to an overall shift by $\Omega/2$, are just the Floquet quasienergies) are given by
\begin{align}\label{dot quasienergy}
\varepsilon_{i}&=\bar E +(-1)^i\sqrt{(\epsilon-\Omega/2)^2+|\Delta|^2},
\end{align}
where $\bar E = (E_1+E_2)/2$ and $\epsilon = (E_1-E_2)/2$.  Observe that the minimum size of the gap is $2|\Delta|$, when $\Omega=2\epsilon$.

\begin{figure}[t]
(a)\includegraphics[width=.475\textwidth]{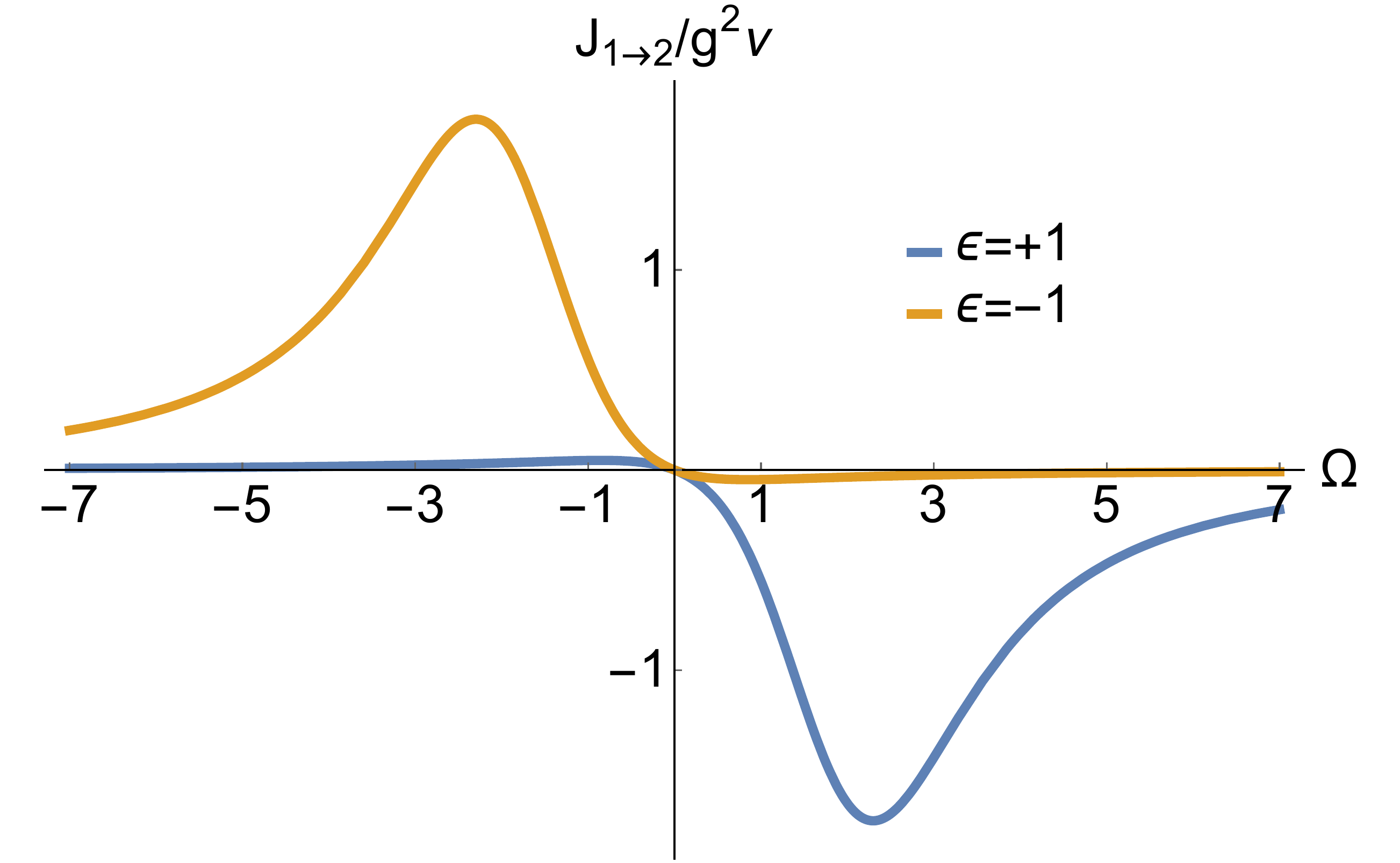}
\\
(b)\includegraphics[width=.475\textwidth]{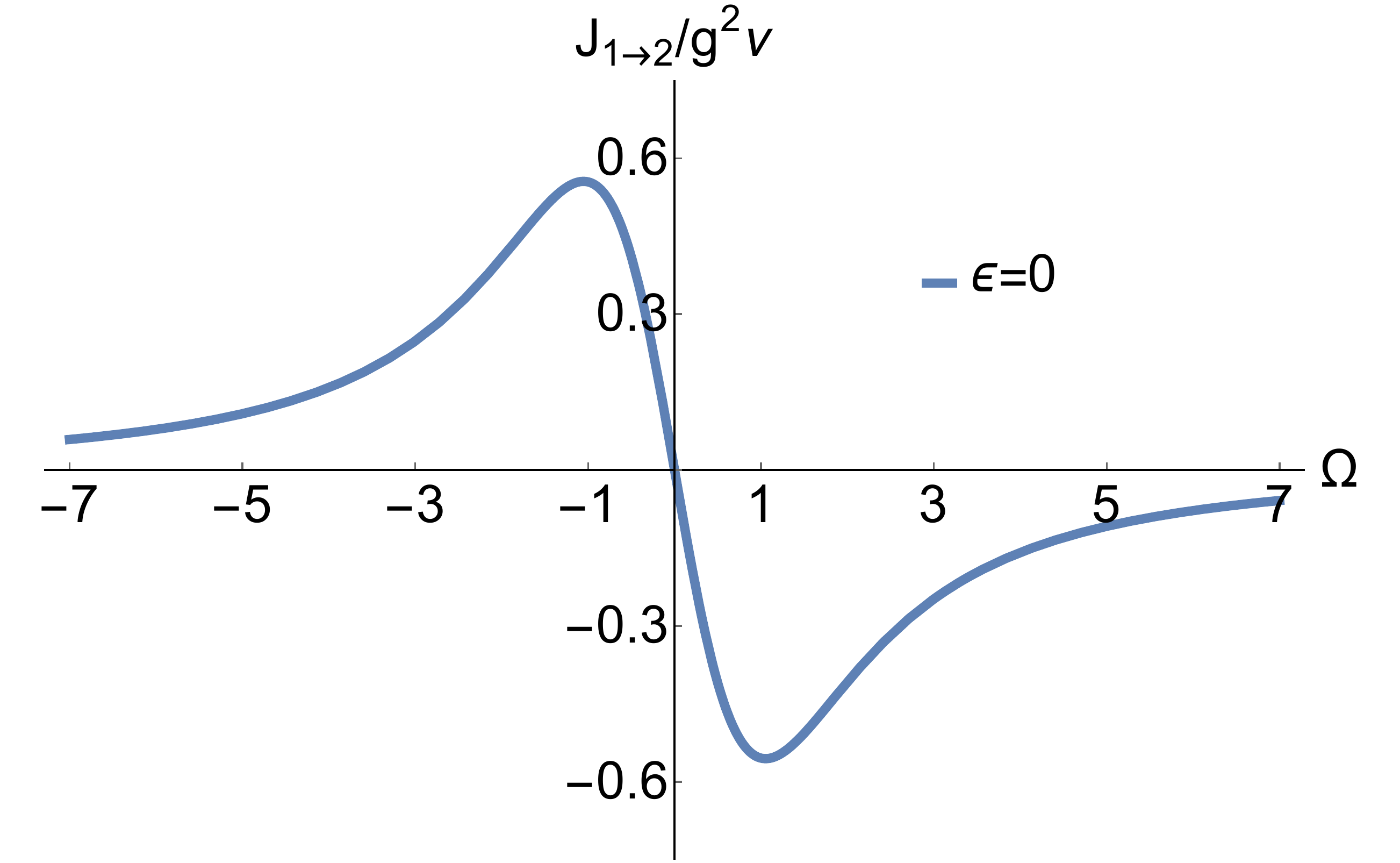}
\caption{(Color online) Persistent electrical current from dot 1 to dot 2 [c.f.~Eq.~\eqref{J1->2}] for (a) nonzero and (b) zero detuning $\epsilon$.  Energies are again measured in units of the rotating-frame gap $2|\Delta|$, and $\beta=2$ as in Fig.~\ref{fig:dot_pops}.  The sign of the current indicates its direction.
\label{fig:dot_current}
        }
\end{figure}

The average steady-state occupation $\tilde n_1$ of dot 1, which is given in Eq.~\eqref{steady-state solution}, is plotted as a function of frequency in Fig.~\ref{fig:dot_pops} for several values of the detuning $\epsilon=(E_1-E_2)/2$.  (The occupation of dot 2 is obtained as $\tilde n_2 = 1-\tilde n_1$, and has therefore been omitted in the Figure.)  We fix the chemical potential in the leads to lie in the middle of the rotating-frame gap, i.e.~$\mu=\bar E$.   Observe that for large $|\Omega|$, the occupations converge to the asymptotic distributions given in Eq.~\eqref{simplest nonthermal populations}, and finally at even larger $|\Omega|$ to the equilibrium values $\tilde n_i=n_F(E_i)$.  This is consistent with the fact that the driving self-averages for large frequencies, as the system does not have time to respond to the rapidly oscillating terms in the Hamiltonian.  Furthermore, note that for $\epsilon\neq 0$ the occupations are asymmetric functions of $\Omega$, becoming most nonthermal for $\Omega\sim\epsilon$.

In this example, the flavor current of Eq.~\eqref{J1->2} is a charge current from dot 1 to dot 2 that flows without a bias between the left and right reservoirs.  In this sense, the driven system acts as a non-equilibrium version of an adiabatic quantum pump.\cite{thouless,brouwer,switkes}  As is evident from Figure~\ref{fig:dot_current}, the current has support precisely where the nonthermal character of the occupations is most pronounced.  For nonzero detuning, the current is also an asymmetric function of $\Omega$, with a resonance near $\Omega=2\epsilon$, where the minimum in the rotating-frame gap occurs.  For $\epsilon=0$, the current is smaller in magnitude, but remains finite (and chiral) despite the lack of detuning between the dots.  This suppression can be attributed to the fact that the driving field is never resonant with any transitions in the system.  Note that in both cases, the direction of the current is set by the sign of $\Omega$, as we will also see in the next example.

\begin{figure}[t]
(a)\includegraphics[width=.475\textwidth]{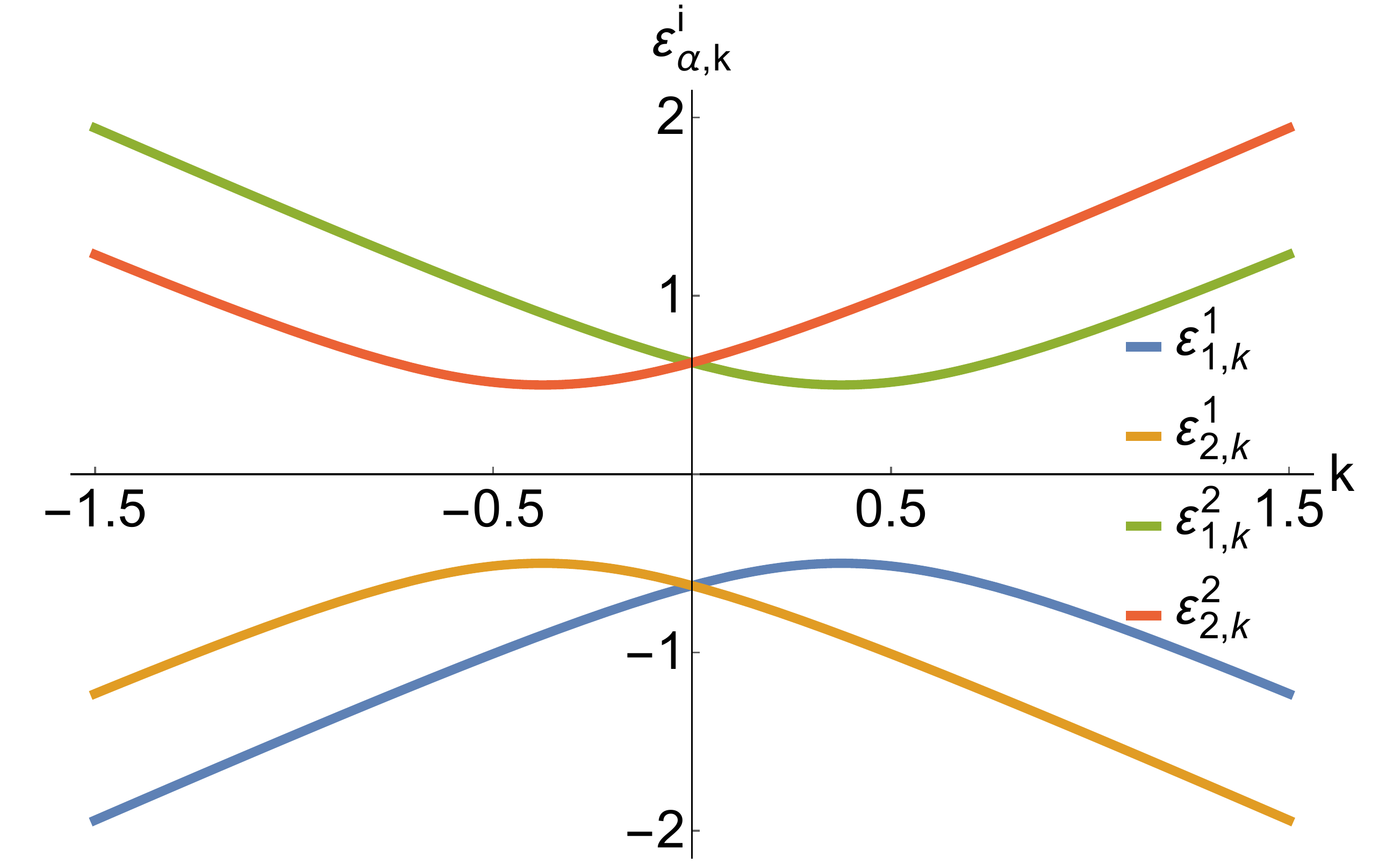}
\\
(b)\includegraphics[width=.475\textwidth]{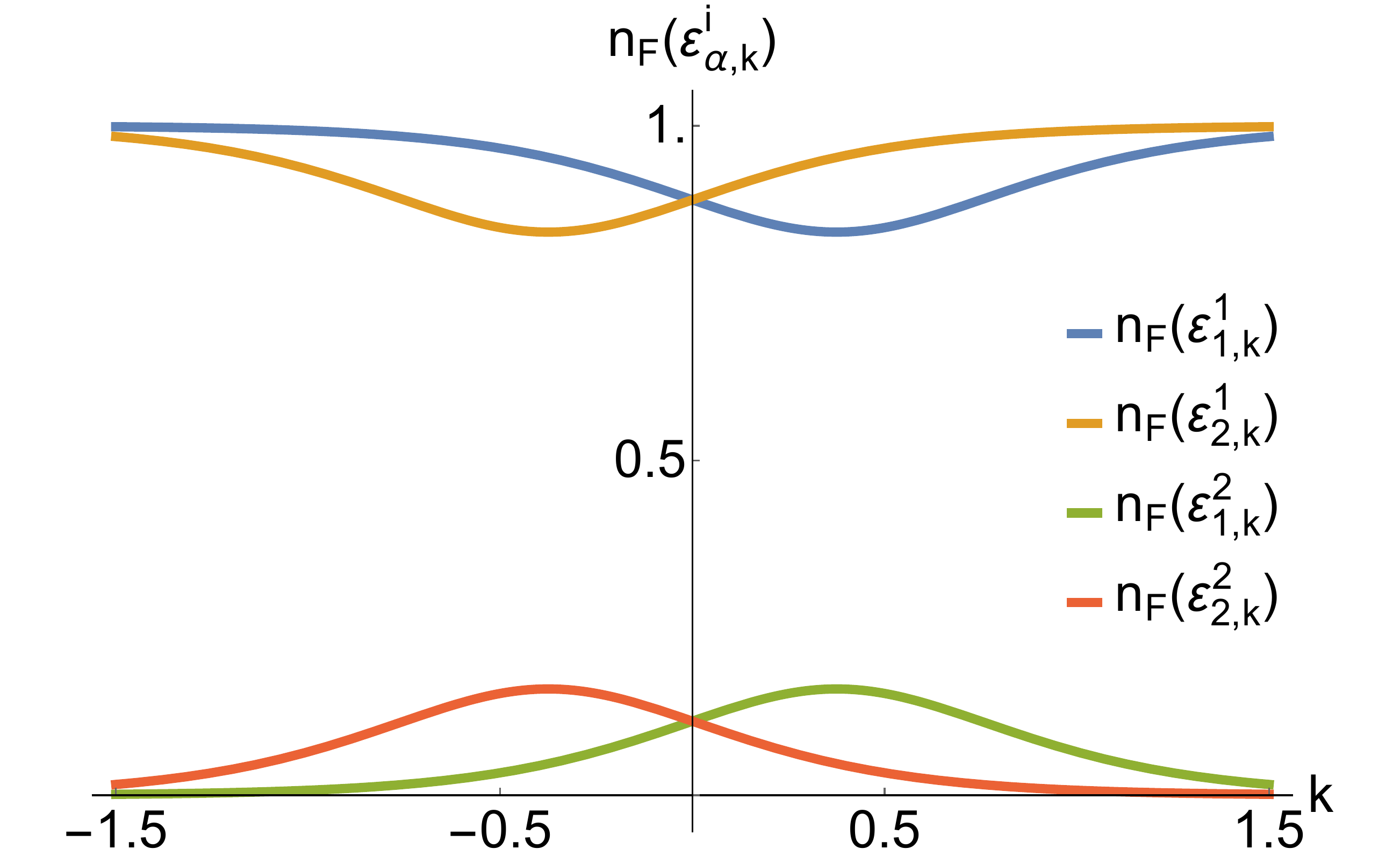}
\\
(c)\includegraphics[width=.475\textwidth]{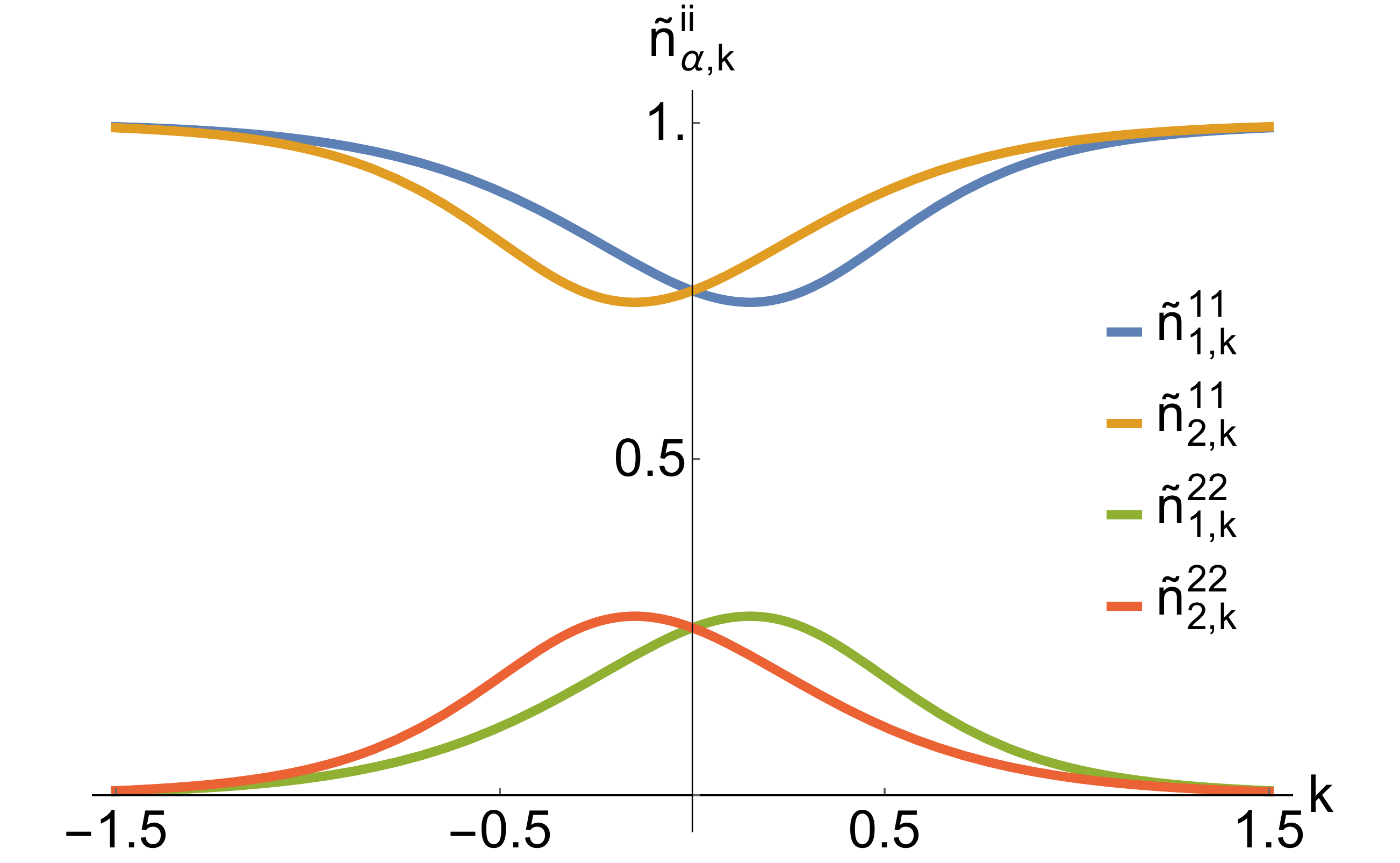}
\caption{(Color online) Nonthermal populations in driven graphene.  The momentum axis represents a cut along the $k_x$-direction, but all quantities are rotationally invariant in the $k_x$-$k_y$ plane.  All energies and momenta are measured in units of the quasienergy gap $2|\Delta|$, and we have chosen an inverse temperature $\beta = 3.3$.  (a) Spectrum of the driven Dirac Hamiltonian in the rotating frame, c.f.~\eqref{rotating graphene spectrum}.  (b) Effective thermal distribution $n_F(\varepsilon^{\, i}_{\alpha,\bm k})$.  (c) Nonthermal quasienergy distribution $\tilde n^{\, ii}_{\alpha,\bm k}$.
\label{fig:graphene_pops}
        }
\end{figure}

\subsubsection{Dirac fermions in driven graphene}\label{graphene example}

As a second example, we consider spinless Dirac fermions in graphene driven by a rotating Kekul\'e mass term,~\cite{hou} which can be induced by the coherent excitation of a particular zone-boundary optical phonon.~\cite{iadecola1}  This model is known to thermalize when the primary dissipation mechanism is acoustic phonons in the graphene flake,~\cite{iadecola2} but we consider here the alternative grand canonical setup discussed in this paper.  We take the fermionic reservoir to be, for example, a substrate that serves as a source and sink of electrons.  Here, the species index $i=1,2$ labels the two inequivalent Dirac cones at momenta $\bm K_\pm$ (see Fig.~\ref{fig:graphene}), and the band index $\alpha=1,2$ labels the positive- and negative-energy branches of the two Dirac cones.  The bare dispersion is given by
\begin{align}
\begin{split}
E^1_{\alpha,\bm k}&=(-1)^{\alpha+1}\, k\\
E^2_{\alpha,\bm k}&=(-1)^\alpha\, k,
\end{split}
\end{align}
where $k=|\bm k|$ is the magnitude of the displacement from either Dirac point, and the spectrum of the driven Hamiltonian in the rotating frame is given by
\begin{align}\label{rotating graphene spectrum}
\varepsilon^{\, i}_{\alpha,\bm k}&= (-1)^i\sqrt{\[k+(-1)^\alpha\, \Omega/2\]^2+|\Delta|^2}.
\end{align}
For any frequency, the quasienergy bands $\varepsilon^{\, i\, \prime}_{\alpha,\bm k}=\varepsilon^{\, i}_{\alpha,\bm k}+\Omega/2$ are separated by a gap $2|\Delta|$, in units of which we measure all energies and momenta in the subsequent discussion.  Also, note that the driving frequency $\Omega$ enters the quasienergy spectrum as a shift in the position of the two Dirac points [see Fig.~\ref{fig:graphene_pops}(a)].  In the discussion below, we choose to set the chemical potential of the reservoir to zero energy, in the middle of the quasienergy gap,~\cite{footnote:quasienergy_shift} so that only the negative-energy states are filled.

In Fig.~\ref{fig:graphene_pops}, we compare the momentum-resolved effective thermal quasienergy distribution $n_F(\varepsilon^{\, i}_{\alpha,\bm k})$ to the steady-state populations $\tilde n^{\, ii}_{\alpha,\bm k}$ in Eqs.~\eqref{steady-state solution}, along with the quasienergy bands given in Eq.~\eqref{rotating graphene spectrum} for reference.  In Fig.~\ref{fig:graphene_pops}(b), we see that the effective thermal distribution shows a depletion in the populations near the top of the valence band, and a corresponding enhancement near the bottom of the conduction band, that is brought about by thermal excitation of the electrons.  The ``hot spots" near which most of the transitions occur are centered at the positions of the shifted Dirac points, $|\bm k|=\Omega/2$, which are the locations of the quasienergy band minima and maxima.  The nonthermal distributions in Fig.~\ref{fig:graphene_pops}(c) are qualitatively similar, but show a transfer of electrons from the valence to the conduction band that is more pronounced than what is seen in the thermal case.  Furthermore, the ``hot spots" where these transitions are most prominent are shifted slightly in momentum space, towards the original positions of the Dirac points at $|\bm k|=0$.

\begin{figure}[t]
\includegraphics[width=.475\textwidth]{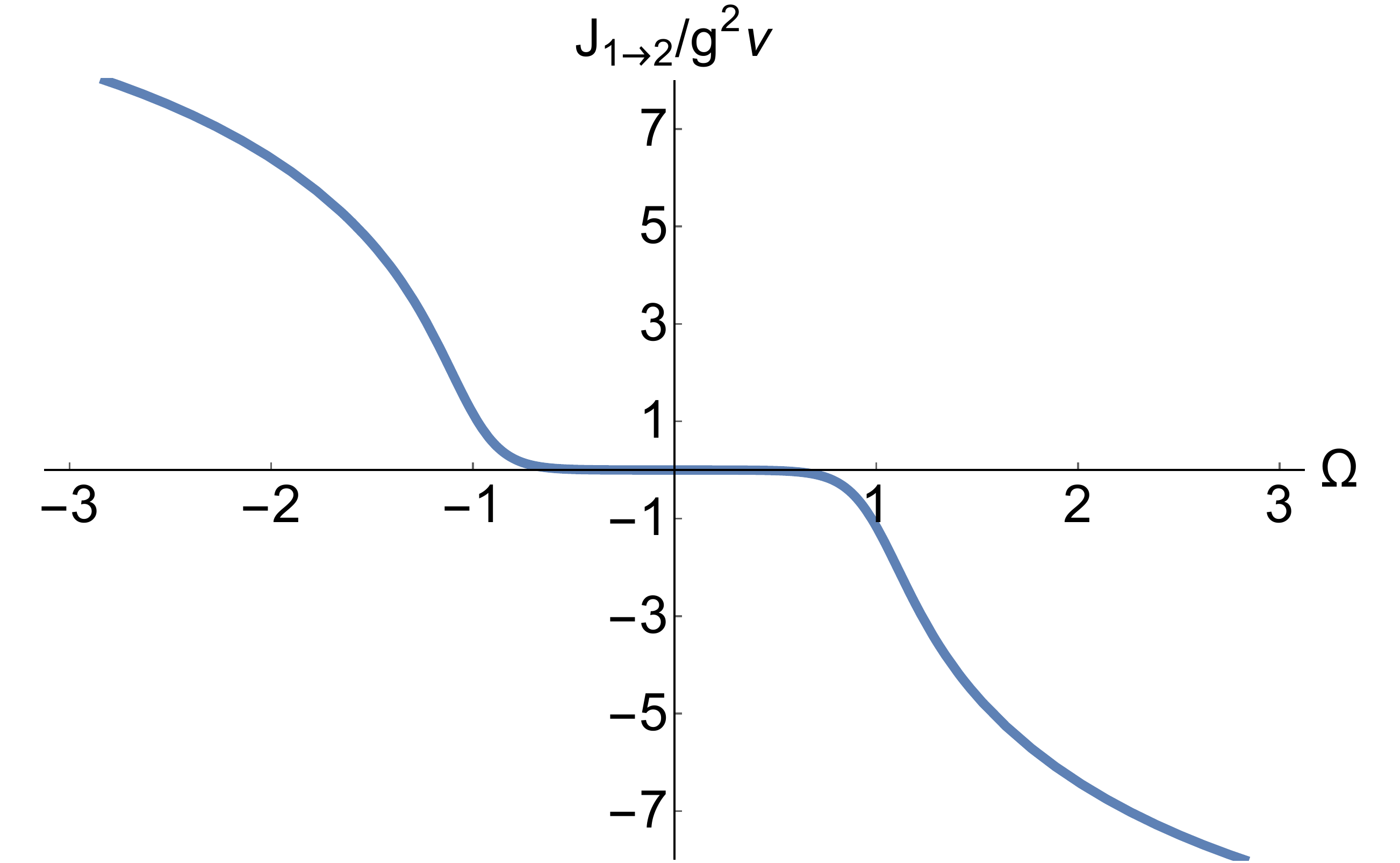}
\caption{(Color online) Persistent valley current in driven graphene.  The suppression of the current for driving frequencies $\Omega<2|\Delta|\equiv 1$ is a consequence of the low temperature ($\beta^{-1}=|\Delta|/10$) used in the plot.  This suppression would be washed out by thermal excitations for $\beta^{-1}\sim |\Delta|$.
\label{fig:graphene_current}
        }
\end{figure}

We now move on to discuss the steady-state current $J_{1\to 2}$, which in our single-particle basis corresponds to a persistent valley current due to scattering between the Dirac points.  This current is plotted as a function of the driving frequency in Fig.~\ref{fig:graphene_current}.  As we saw in the case of the double-dot at zero detuning, $J_{1\to2}$ is an odd function of the driving frequency, indicating a change in direction of the scattering when $\Omega$ changes sign.  Furthermore, at temperatures far below the quasienergy gap, there is a very clear activation barrier for the current, which flows only when $\Omega>2|\Delta|$, up to thermal broadening.  Such a barrier is also present in the current from the double-dot example (c.f.~Fig.~\ref{fig:dot_current}) at sufficiently low temperatures $\beta^{-1}\ll 2|\Delta|$.  At large frequencies, the current in Fig.~\ref{fig:graphene_current} does not taper off as it does in Fig.~\ref{fig:dot_current}, as the rotating-frame spectrum is not bounded from above or below [c.f.~Eq.~\eqref{rotating graphene spectrum}] and therefore remains resonant with some driving frequency $\Omega$ for all $k$.  This is purely an artifact of our approximation of the driven graphene Hamiltonian by a Dirac-type Hamiltonian---a treatment of the problem at the level of the time-dependent tight-binding Hamiltonian would provide a natural frequency cutoff due to the finite bandwidth.  The expected suppression of the current for $\Omega$ much larger than the bandwidth would then become apparent.

\section{Conclusion}
In this paper, we have studied the possibility of effective thermalization in Floquet systems coupled to reservoirs with which they can exchange both energy and particles.  We found that, while thermalization is possible, it does not happen in most cases of interest as it requires a sufficiently simple driving protocol and fine-tuning of the system-bath coupling.  We illustrated these ideas by studying an example of such a system that can either effectively thermalize with fine tuning or not thermalize at all.  Using a Born-Markov master equation approach, we calculated analytically the populations and coherences in the steady state of this system, and found that, for arbitrary system-bath couplings away from the critical line $g_2=0$, the system relaxes to a nonthermal quasienergy distribution with a persistent current at zero external bias.  We then studied this steady-state solution for two physical examples, and observed the nontrivial behavior of the populations and persistent currents as functions of frequency.

There remain many promising directions for future work on the subject of nonthermal steady states in solid-state systems.  First, since the bare dispersions $E^{\, i}_{\alpha,\bm k}$ were arbitrary, it would be interesting to study the class of models treated in this paper for more complicated driven lattice models than the graphene example considered here.  This would highlight the role of dispersion in shaping the steady-state currents and Floquet occupations.  Second, it would be interesting to determine the role of the bath density of states in the steady-state distribution, and to examine to what extent the bath density of states could be manipulated to design the steady-state populations and currents.  Third, the fate of periodically-driven interacting systems when coupled to thermal reservoirs is largely unexplored.  Several recent works\cite{d'alessio_polkovnikov,d'alessio_rigol,ponte,lazarides2} have predicted that closed interacting systems heat up to infinite temperature when driven periodically.  It would be interesting to explore the landscape of possible interacting steady states that can be accessed when this heating is balanced by dissipation to the reservoir.  In studying such systems, one pursues the much deeper problem of achieving phases of matter via nonthermal steady states that are truly inaccessible at equilibrium.

\acknowledgments{We thank Camille Aron, Garry Goldstein, and Tadeusz Pudlik for helpful discussions.  We are also especially grateful to Luca D'Alessio, who shared with us preliminary results regarding lack of effective thermalization in a periodically-driven system where particles are not exchanged with the reservoir.~\cite{d'alessio}  T.I. was supported by a National Science Foundation Graduate Research Fellowship under Grant No. DGE-1247312, and C.C. was supported by DOE Grant DEF-06ER46316.}

\appendix*
\section{Expressions for coefficients in Eq.~\eqref{EOM populations decoupled}}
In this Appendix we record expressions for the coefficients entering the equations of motion for the populations and coherences.  They are
\begin{subequations}
\begin{align}
|\tilde g^{11}_{\alpha,\bm k}|^2\, \nu^{11}_{\alpha,\bm k}&=\frac{1}{2}\, g^2\,  \nu^{11}_{\alpha,\bm k} \left[1- \frac{\(\epsilon_{\alpha,\bm k}-\frac{\Omega}{2} \)}{\delta E_{\alpha,\bm k}}\right]\\
|\tilde g^{12}_{\alpha,\bm k}|^2\, \nu^{12}_{\alpha,\bm k}&=\frac{|\Delta|^2\,  g^2\,  \nu^{11} _{\alpha,\bm k}}{|\Delta|^2+\left[\(\epsilon_{\alpha,\bm k}-\frac{\Omega}{2}\)-\delta E_{\alpha,\bm k}\right]^2}\\
\tilde g^{12}_{\alpha,\bm k}\,\tilde g^{22\, *}_{\alpha,\bm k}\, \nu^{22}_{\alpha,\bm k}&=\frac{|\Delta|\,  g^2\,  \nu^{22} _{\alpha,\bm k}}{2\, \delta E_{\alpha,\bm k}}\\
\tilde g^{11}_{\alpha,\bm k}\,\tilde g^{21\, *}_{\alpha,\bm k}\, \nu^{21}_{\alpha,\bm k}&=-\frac{|\Delta|\,   g^2\,  \nu^{22} _{\alpha,\bm k}}{2\, \delta E_{\alpha,\bm k}}\\
|\tilde g^{22}_{\alpha,\bm k}|^2\, \nu^{22}_{\alpha,\bm k}&=\frac{|\Delta|^2\,  g^2\,  \nu^{22} _{\alpha,\bm k}}{|\Delta|^2+\left[\(\epsilon_{\alpha,\bm k}-\frac{\Omega}{2}\)+\delta E_{\alpha,\bm k}\right]^2}\\
|\tilde g^{21}_{\alpha,\bm k}|^2\, \nu^{21}_{\alpha,\bm k}&=\frac{1}{2}\, g^2\,  \nu^{22} _{\alpha,\bm k}\left[1+ \frac{\(\epsilon_{\alpha,\bm k}-\frac{\Omega}{2} \)}{\delta E_{\alpha,\bm k}}\right]\\
\tilde g^{21\, *}_{\alpha,\bm k}\,\tilde g^{11}_{\alpha,\bm k}\, \nu^{11}_{\alpha,\bm k}&=-\frac{|\Delta|\,  g^2\,  \nu^{11} _{\alpha,\bm k}}{2\, \delta E_{\alpha,\bm k}}\\
\tilde g^{22\, *}_{\alpha,\bm k}\,\tilde g^{12}_{\alpha,\bm k}\, \nu^{12}_{\alpha,\bm k}&=\frac{|\Delta|\,   g^2\,  \nu^{11} _{\alpha,\bm k}}{2\, \delta E_{\alpha,\bm k}}.
\end{align}
\end{subequations}
Note that above we have made use of the assumptions that $g_1=g_2=g$ and $\nu^{\, i1}_{\alpha,\bm k}=\nu^{\, i2}_{\alpha,\bm k}$.

In the main text, we also make use of high-frequency limits of the above expressions.  To order $1/|\Omega|$, they are
\begin{subequations}
\begin{align}
|\tilde g^{11}_{\alpha,\bm k}|^2\, \nu^{11}_{\alpha,\bm k} &\approx \frac{1}{2}\, g^2\, \nu^{11}_{\alpha,\bm k}\(1+\text{sgn }\Omega\)\\
|\tilde g^{12}_{\alpha,\bm k}|^2\, \nu^{12}_{\alpha,\bm k} &\approx \frac{1}{2}\, g^2\, \nu^{11}_{\alpha,\bm k}\(1-\text{sgn }\Omega\)\\
|\tilde g^{22}_{\alpha,\bm k}|^2\, \nu^{22}_{\alpha,\bm k} &\approx \frac{1}{2}\, g^2\, \nu^{22}_{\alpha,\bm k}\(1+\text{sgn }\Omega\)\\
|\tilde g^{21}_{\alpha,\bm k}|^2\, \nu^{21}_{\alpha,\bm k} &\approx \frac{1}{2}\, g^2\, \nu^{22}_{\alpha,\bm k}\(1-\text{sgn }\Omega\)\\
\tilde g^{12}_{\alpha,\bm k}\,\tilde g^{22\, *}_{\alpha,\bm k}\, \nu^{22}_{\alpha,\bm k} &\approx g^2 \abs{\frac{\Delta}{\Omega}}\, \nu^{22}_{\alpha,\bm k}\\
\tilde g^{21\, *}_{\alpha,\bm k}\,\tilde g^{11}_{\alpha,\bm k}\, \nu^{11}_{\alpha,\bm k} &\approx -g^2 \abs{\frac{\Delta}{\Omega}}\, \nu^{11}_{\alpha,\bm k}.
\end{align}
\end{subequations}
These limits of the coefficients were used to derive Eq.~\eqref{simplest nonthermal populations}.

\bibliographystyle{apsrev}

\bibliography{refs_grand_Floquet}

\begin{thebibliography}{48}
\bibitem[{foo({\natexlab{a}})}]{footnote:cuetara}
\bibinfo{howpublished}{It is worth pointing out that this result is in
  agreement with the findings of Ref.~\onlinecite{cuetara}, which studied an
  instance of this class of problems in the high-frequency limit.}
\bibitem[{foo({\natexlab{b}})}]{footnote:quasienergy_shift}
\bibinfo{howpublished}{Strictly speaking, since the Floquet quasienergies are
  given by $\varepsilon^{\, i}_{\alpha,\bm k}+\Omega/2$, one should set
  $\mu=\Omega/2$ in order to keep the chemical potential inside the quasienergy
  gap for all $\Omega$.}
\end{thebibliography}

\end{document}